%% file: ms.tex
\documentclass[bg, manuscript]{copernicus}

\usepackage{longtable}
\usepackage{booktabs}

\begin{document}
\nolinenumbers
\title{Representation of the Terrestrial Carbon Cycle in CMIP6}

% \Author[affil]{given_name}{surname}

\Author[1,2]{Bettina~K.}{Gier}
\Author[2]{Manuel}{Schlund}
\Author[3,4]{Pierre}{Friedlingstein}
\Author[5]{Chris~D.}{Jones}
\Author[6]{Colin}{Jones}
\Author[7]{S\"{o}nke}{Zaehle}
\Author[2,1]{Veronika}{Eyring}

\affil[1]{University of Bremen, Institute of Environmental Physics (IUP), Bremen, Germany}
\affil[2]{Deutsches Zentrum f\"{u}r Luft- und Raumfahrt (DLR), Institut f\"{u}r Physik der Atmosph\"{a}re, Oberpfaffenhofen, Germany}
\affil[3]{College of Engineering, Mathematics and Physical Sciences, University
of Exeter, Exeter, EX4 4QE, United Kingdom}
\affil[4]{LMD/IPSL, ENS, PSL Universit\'{e}, \'{E}cole Polytechnique, Institut
Polytechnique de Paris,\hack{\break} Sorbonne Universit\'{e}, CNRS, Paris, France}
\affil[5]{Met Office Hadley Centre, Exeter, UK}
\affil[6]{National Centre for Atmospheric Science, University of Leeds, UK}
\affil[7]{Biogeochemical Signals Department, Max Planck Institute for Biogeochemistry, Jena, Germany}

%% The [] brackets identify the author with the corresponding affiliation. 1, 2, 3, etc. should be inserted.

%% If an author is deceased, please mark the respective author name(s) with a dagger, e.g. "\Author[2,$\dag$]{Anton}{Smith}", and add a further "\affil[$\dag$]{deceased, 1 July 2019}".

%% If authors contributed equally, please mark the respective author names with an asterisk, e.g. "\Author[2,*]{Anton}{Smith}" and "\Author[3,*]{Bradley}{Miller}" and add a further affiliation: "\affil[*]{These authors contributed equally to this work.}".

\correspondence{Bettina K. Gier (gier@uni-bremen.de)}

\runningtitle{Representation of the Terrestrial Carbon Cycle in CMIP6}
\runningauthor{B.~K.~Gier et al.}

\received{}
\pubdiscuss{} %% only important for two-stage journals
\revised{}
\accepted{}
\published{}

%% These dates will be inserted by Copernicus Publications during the typesetting process.

\firstpage{1}

\maketitle

\begin{abstract}
Improvements in the representation of the land carbon cycle in Earth system models participating in the Coupled Model Intercomparison Project Phase 6 (CMIP6) include interactive treatment of both the carbon and nitrogen cycles, improved photosynthesis, and soil hydrology. To assess the impact of these model developments on aspects of the global carbon cycle, the Earth System Model Evaluation Tool (ESMValTool) is expanded to compare CO$_{2}$ concentration and emission driven historical simulations from CMIP5 and CMIP6 to observational data sets. A particular focus is on the differences in models with and without an interactive terrestrial nitrogen cycle. Overestimations of photosynthesis (gross primary productivity (GPP)) in CMIP5 were largely resolved in CMIP6 for participating models with an interactive nitrogen cycle, but remaining for models without one. This points to the importance of including nutrient limitation. Simulating the leaf area index (LAI) remains challenging with a large model spread in both CMIP5 and CMIP6. In ESMs, global mean land carbon uptake (net biome productivity (NBP)) is well reproduced in the CMIP5 and CMIP6 multi-model means. However, this is the result of an underestimation of NBP in the northern hemisphere, which is compensated by an overestimation in the southern hemisphere and the tropics. Carbon stocks remain a large uncertainty in the models. While vegetation carbon content is slightly better represented in CMIP6, the inter-model range of soil carbon content remains the same between CMIP5 and CMIP6. Overall, a slight improvement in the simulation of land carbon cycle parameters is found in CMIP6 compared to CMIP5, but with many biases remaining, further improvements of models in particular for LAI and NBP is required. Models from modeling groups participating in both CMIP phases generally perform similarly or better in their CMIP6 compared to their CMIP5 models. This improvement is not as significant in the multi-model means due to more new models in CMIP6, especially those using older versions of the Community Land Model (CLM). Emission driven simulations perform just as well as concentration driven models despite the added process-realism. Due to this we recommend ESMs in future CMIP phases to perform emission driven simulations as the standard so that climate-carbon cycle feedbacks are fully active. The inclusion of nitrogen limitation led to a large improvement in photosynthesis compared to models not including this process, suggesting the need to view the nitrogen cycle as a necessary part of all future carbon cycle models. Possible benefits when including further limiting nutrients such as phosphorus should also be considered. 
\end{abstract}

% \copyrightstatement{TEXT} %% This section is optional and can be used for copyright transfers.

\introduction  %% \introduction[modified heading if necessary]
Earth System Models (ESMs) simulate the climate system by interactively coupling physical general circulation models of the atmosphere, ocean, land and cryosphere with biogeochemical and biophysical cycles \citep{Jones2020}. The Coupled Model Intercomparison Project \citep[CMIP,][]{Meehl2000} was established to facilitate a consistent comparison between different ESMs through the use of common forcings and a uniform output structure in order to better understand past, present and future climate. The newest phase, CMIP6 \citep{EyringCMIP6}, provides a large ensemble of model simulations and includes 23 CMIP6-Endorsed Model Intercomparison Projects (MIPs) which facilitate a better analysis of specific scientific questions. Every new phase of CMIP sees additional models and improved model components, making a comparison to previous phases vital to determine if known systematic biases have been reduced and model weaknesses were identified and overcome \citep{Eyring2019}, or if increased model realism through the inclusion of additional processes introduces new biases. Increasing the process-realism of models, for example by replacing time-invariant observation-based fields with interactive prognostic ones, while having a neutral impact on the present day performance of the model can be viewed as a successful step in model improvement. This is particularly important for improved climate projections, even at the possible cost of worse present-day evaluation. \citet{Bock2020} assessed atmospheric variables of the CMIP6 ensemble and compared them to CMIP3 and CMIP5 output. They find that for temperature, precipitation, water vapor and zonal wind speed many long-standing biases remain in the multi-model mean, but individual models and high-resolution versions of models show some improvements in many long-standing biases for temperature and precipitation. In this study, we expand this assessment to the carbon cycle.

The carbon cycle is an important part of ESMs due to the key role of anthropogenic emissions of carbon dioxide (CO$_2$) in driving climate change \citep{IPCC2021}. The land carbon cycle dominated the uncertainty of the global carbon cycle \citep{ipcc6ch5}. The Intergovernmental Panel on Climate Change (IPCC) Fifth Assessment Report (AR5) was largely supported by model simulations from the CMIP5 ensemble, and CMIP6 models were an important input to AR6 \citep{IPCC2021}. It is important to evaluate these models and document their changes compared to CMIP5. The focus of this study is on the results of the CMIP6 historical simulations, split into simulations with prescribed greenhouse gas concentrations and those with prescribed CO$_2$ emissions. Simulations driven by carbon emissions require an interactive carbon cycle to determine the distribution of natural and anthropogenic carbon fluxes across the land, marine and atmospheric reservoirs instead of relying on prescribed atmospheric CO$_2$ concentrations \citep{Friedlingstein2014}. Therefore, only emission driven simulations have fully active climate-carbon cycle feedbacks vital for self-consistent future projections. \citet{Sanderson2023} thus petition for the use of emission driven simulations to be prioritized in CMIP7. This leads to further importance of the evaluation of the carbon cycle as more models will be required to implement and improve their interactive carbon cycle. An analysis of idealized 1\% CO$_2$ increase per year (1pctCO$_2$) simulations for CMIP5 and CMIP6 models with respect to carbon-concentration and carbon-climate feedbacks has been carried out by \citet{Arora2020}. They found that while these feedback parameters have not changed significantly between CMIP5 and CMIP6, the land feedback parameters are weaker for models including a nitrogen cycle coupled to the carbon cycle. \citet{Davies-Barnard2020} documented the development and inclusion of modelling of the terrestrial nitrogen cycle in ESM land-surface schemes, and show how this affects the response to elevated CO$_2$ across models. They find that these models show a more accurate response in the tropics than high-latitudes compared with observed responses. \citet{Gier2020} investigated the atmospheric CO$_2$ concentrations for emission driven CMIP5 and CMIP6 models and found that while CMIP6 models show an improvement in reproducing the observations compared to CMIP5, during the period of the satellite observations (2003-2014) the growth rate is overestimated and the seasonal cycle amplitude is underestimated in both CMIP5 and CMIP6 relative to observations. Furthermore, the model spread in simulated atmospheric CO$_2$ ($\sim$ 45\,ppmv in CMIP5, $\sim$ 35\,ppmv in CMIP6) was found to remain many times larger than the observational uncertainty of under 1\,ppmv over this period.

\citet{Anav2013} investigated the land and ocean carbon cycle for CMIP5 historical model simulations. While most models reproduced the main climatic variables and their seasonal evolution correctly, weaknesses were found in their ability to reproduce more specific biogeochemical fields such as a general overestimation of photosynthesis. Consequently, these were some of the main areas tackled in updating and enhancing the land carbon models of the ESMs for CMIP6, including the addition of a coupled nitrogen cycle and nitrogen limitation which can limit the rates of carbon cycling through vegetation and soil, as well as photosynthesis updates and soil hydrology improvements \citep[e.g.,][]{Danabasoglu2020, Delire2020, Wiltshire2021}. This paper assesses the impacts of these model improvements and additions, especially the impact of an additional coupled nitrogen cycle, and whether they help overcome some of the weaknesses identified in CMIP5. Expanding on the analysis from \citet{Anav2013} for CMIP5, this study uses CMIP6 concentration and emission driven historical simulations to compare to the land carbon cycle in CMIP5 and identify possible improvements in the newer model generation. \citet{Jones2023} drew on expert assessment of regional carbon budgets to evaluate the terrestrial carbon cycle in CMIP6 models and came to similar conclusions. The multi-model mean performs well in most regions for most variables (both carbon fluxes and stocks), but individual models have strengths and weaknesses. In this study we restrict ESM evaluation to datasets with global extent.

Section 2 describes the data used in this study, while Section 3 contains the analysis of different carbon cycle variables in the comparison between CMIP ensembles and observations. Long-term trends and the seasonal cycle of carbon cycle variables are considered, and the analysis is concluded with a performance metrics plot evaluating the climatological seasonal cycle of the models with different observational data sets. Section 4 summarizes the results and conclusions.

\section{Data and Tools}

\subsection{CMIP simulations}
Model simulations from both CMIP Phase 6 \citep{EyringCMIP6}, and Phase 5 \citep{Taylor2012} are used, with Tables \ref{tab:CMIP6} and \ref{tab:CMIP5} listing model characteristics such as their atmosphere and land model components in addition to their main references. A more comprehensive summary of the land model components of the CMIP models is given in Appendix A. Models were selected due to their availability on Earth System Grid Formation (ESGF) nodes for the considered variables.

This study focuses on historical simulations, which aim to reproduce the observed climate since the pre-industrial times. They span from 1850 to 2005 (CMIP5) and 1850 to 2014 (CMIP6). Both simulations with prescribed greenhouse gas concentrations (concentration driven) and prescribed CO$_2$ emissions (emission driven) are considered, but evaluated separately and compared to each other. Models participating in the emission driven simulations, marked in bold in the tables, use their interactive carbon cycle to determine the distribution of natural and anthropogenic carbon fluxes across the land, marine and atmospheric reservoirs instead of relying on prescribed atmospheric CO$_2$ concentrations \citep{Friedlingstein2014}.

Very few CMIP5 models had a coupled carbon-nitrogen cycle. While the BNU-ESM model included carbon-nitrogen interactions, they were turned off for the CMIP5 model simulations as the nitrogen cycle had not been fully evaluated \citep{Ji2014}. Therefore, a nitrogen cycle was included in two out of 18 CMIP5 models (CESM1-BGC, NorESM1-ME) which both use the CLM4 land model and in 15 out of 23 CMIP6 models spread over six different land models - with CLM in different versions accounting for eight CMIP6 models (v5: CESM2, CESM2-WACCM, NorESM2-LM, NorESM2-MM, v4.5: CMCC-CM2-SR5, CMCC-ESM2, v4: SAM0-UNICON, TaiESM1). The other land models in CMIP6 with a coupled nitrogen cycle are LPJ-GUESS (EC-Earth3-CC, EC-Earth3-Veg), JSBACH (MPI-ESM-1-2-HAM, MPI-ESM1-2-LR), CABLE+CASA-CNP (ACCESS-ESM1-5), JULES-ES (UKESM1-0-LL) and Visit-e (MIROC-ES2L). This shows a large bias towards the CLM land model in CMIP6 which needs to be considered while analyzing the multi-model mean (MMM).

To facilitate a direct comparison of CMIP5 and CMIP6 data in figures containing temporal means, only data up to 2005 representing the end of the CMIP5 historical simulations is considered. Unless stated otherwise, figures use mean data over the time period 1986-2005. Only one realization per model is used, as different ensemble members perform similarly to each other with respect to the carbon cycle and using an ensemble mean would lead to an under representation of the internal variability present in individual ensemble members. Multi-model means (MMMs) were computed separately for each project and experiment combination, as well as an additional distinction between models with and without interactive nitrogen models, and are computed on the monthly gridded data, for which models are regridded to a common 2$^\circ$x2$^\circ$ grid. MMMs are neither weighted according to the interdependence of the models and model components, nor according to their performance relative to observational products.

While we split models into groups only dependent on the presence of an interactive nitrogen cycle in this study, vegetation dynamics is another important process for ESM comparison. Models interactively simulating vegetation cover may simulate trees or grasses in the wrong areas compared to models using observational land cover maps, impacting variables with a strong relation to land cover, such as LAI or GPP. While models with prescribed land cover may show better LAI in the present day, they cannot predict future changes in vegetation cover nor their impact on regional climate and carbon processes. For reference, Tables \ref{tab:CMIP6} and \ref{tab:CMIP5} note models with dynamic vegetation with a D in the comment column.

\subsection{Reference Data}
A large range of observations and reanalysis data sets have been used to assess model performance. These data sets are listed in Table 3 along with their main reference(s), their source, the variables used and their temporal coverage. Both observational and reanalysis data sets will be referred to as “observations” from here on, to contrast the results from the CMIP model simulations. The longest observational records are derived from reanalyses, while satellite observations only provide data since the late 20th century. Since most reference data sets do not come with observational uncertainty, a common approach is to use several reference data sets per variable where available, as noted in \citet{Seiler2022}. This approach is also taken in this study.

For the leaf area index (LAI), we use the LAI3g product \citep{Zhu2013} that provides global monthly gridded data starting in the year 1981. It has been generated using an artificial neural network based on data from the Advanced Very High Resolution Radiometer (AVHRR) and the Moderate Resolution Imaging Spectroradiometer (MODIS). Furthermore, we also use the newly released GIMMS LAI4g data set \citep{Cao2023_lai4g} which is based on the same satellite data as LAI3g but employs a newer NDVI data set base which removes the effects of the satellite orbital drift and AVHRR sensor degradation, which plagued many other LAI data sets. Furthermore, LAI4g uses a large number of high-quality Landsat LAI samples to increase the spatiotemporal consistency of the data set. Lastly, the Global Land Surface Satellite \citep[GLASS; ][]{Liang2021_GLASS} is a product suite with 12 products, of which we employ both the leaf area index (LAI) and gross primary productivity (GPP) products. Similarly to LAI3g and LAI4g, GLASS is based on data obtained from AVHRR and MODIS. As newer GLASS data products only use MODIS and thus start from 2000, this paper uses a previous GLASS version (v4.0) which includes AVHRR data and thus starts in 1981 for LAI and 1982 for GPP respectively. GLASS LAI uses general regression neural networks trained on preprocessed reflectance data of an entire year to estimate the one-year LAI profile for each pixel. The LAI product is one of the variables used to estimate GLASS GPP with an Eddy Covariance-Light Use Efficiency model. Both GLASS products are available on a 0.05° grid with a frequency of 8 days.

Another GPP product, MTE \citep{Jung2011}, provides global monthly gridded data starting in 1982. It uses an upscaling of data from the FLUXNET eddy covariance tower network based on the model tree ensembles (MTE) approach. Similarly, the FLUXCOM product \citep{Jung2019} is also based on an upscaling of FLUXNET site level observations, but additionally incorporates a larger variety of machine learning methods, and also includes remote sensing (from MODIS) and meteorological data. Here, we use a global monthly gridded version of FLUXCOM (starting in 1980) from the RS+METEO setup. Due to the assumption of an unchanging average CO$_2$ level, both MTE and FLUXCOM data is known to have an unrealistic non-existent trend (0.01\,PgC yr$^{-2}$ globally) \citep{Anav2015}. Thus trend analysis on GPP should exclude these data sets.

The main data set for the land-atmosphere carbon flux (NBP) is the JENA-CarboScope (version sEXTocNEET\_v2020) product \citep{Roedenbeck2005}, which provides global daily gridded data starting from the year 1957. This data set provides surface-atmosphere CO$_2$ fluxes based on atmospheric measurements calculated from an atmospheric transport inversion. The inversion used here (NEE-T inversion) involves a regression of interannual net ecosystem exchange (NEE) anomalies against air temperature anomalies (T). In total, JENA-CarboScope uses data from 156 atmospheric measurement sites distributed across the entire globe. The alternative data set for the land-atmosphere carbon flux is a further inversion product from the Copernicus Atmosphere Monitoring Service \citep[CAMS;][]{Chevallier2005, Chevallier2010, Chevallier2013}. CAMS provides global gridded data on a monthly resolution starting in 1979 (other temporal resolutions are also available). The inversion product we use here (v20r2) is based on surface measurements from more than 100 sites. A third data set used for comparing the global annual mean NBP is the Global Carbon Project \citep[GCP;][]{Friedlingstein2022}, which estimates the global carbon budget using several observations and models. It provides estimates for emissions from fossil fuel combustion and industrial processes, emissions from land-use change, atmospheric CO$_2$ growth rate, ocean sink, land sink, cement carbonation sink and the budget imbalance from combining all these terms. The land-atmosphere carbon flux for GCP has to be calculated by subtracting the land-use change emissions from the residual land sink. The land sink was obtained from averaging the results from seventeen Dynamic Global Vegetation Models (DGVMs) which reproduce the observed mean total land uptake of the 1990s and is given with an uncertainty of $\pm$ 0.5\,PgC yr$^{-1}$ on average. The land use change emissions are estimated from the average of three bookkeeping models with an uncertainty of $\pm$ 0.7\,PgC yr$^{-1}$, making it one of the only data sets with direct estimations for uncertainties.

For each of the remaining carbon cycle variables, only a single reference data set is taken into account.  For the vegetation and soil carbon pools, the NDP-017b \citep{Gibbs2006} and HWSD+NCSCD \citep{Wieder2014, Hugelius2013_NCSCD}  products are used, respectively. Both data sets provide global gridded annual data for the single year 2000. NDP-017b uses an updated database that extends the methodology of \citet{Olson1985}, who developed a global carbon stocks map of above and below ground biomass using 20 years of field investigations, consultations and literature analysis, to more up-to-data land cover conditions of the Global Land Cover Database (GLC2000). The Harmonized World Soil Database (HWSD) uses large volumes of regional and national soil information to create an empirical data set that provides soil parameter estimates for topsoil (0–30 cm) and subsoil (30–100 cm). Similar to \citet{Varney2022} we combine the HWSD data set with the Northern Circumpolar Soil Carbon Database \citep[NCSCD; ][]{Hugelius2013_NCSCD} to complement the HWSD data in the polar region. It uses data on soil order coverage to calculate soil organic carbon content and mass with 1778 pedon data. Wherever overlap between the two data sets occurs, the NCSCD data is chosen.

\subsection{ESMValTool}
The analysis in this paper was produced using the Earth System Model Evaluation Tool (ESMValTool) version 2 \citep{Righi2020, Eyring2020, Lauer2020, Weigel2021, Schlund2023}. ESMValTool is an open source community diagnostics and performance metrics tool which has been developed to routinely evaluate ESMs contributing to CMIP and compare them with other ESMs, predecessor versions, as well as observations. Since its first release \citep{Eyring2016}, ESMValTool has been updated for increased performance in its core functionality to deal with the increased data volume of CMIP6, and now features full traceability and reproducibility through provenance, as well as new and updated diagnostics and metrics, which can be applied to many models and variables. Available diagnostics cover a large range of scientific topics and are described in three papers. They include large-scale diagnostics for quasi-operational and comprehensive evaluation of ESMs \citep{Eyring2020}, diagnostics for extreme events, regional model and impact evaluation and analysis \citep{Weigel2021}, and diagnostics for emergent constraints and analysis of future projections \citep{Lauer2020}. A new suite of recipes has been developed covering the work of this study, as well as some improvements on previous diagnostics for the carbon cycle available in ESMValTool. This facilitates the evaluation of the carbon cycle in future studies, including the analysis of upcoming CMIP7 simulations that can be easily compared to CMIP5 and CMIP6 to assess improvements.

\section{CMIP model performance}

General climate variables, such as temperature and precipitation have a large influence on the carbon cycle. It is therefore important to assess how well these variables are simulated by the ESMs. If they are well reproduced but carbon cycle variables are not, it is likely due to a poor representation of processes specific to the carbon cycle, while a poor performance in the physical variables makes an attribution of the cause of poor performance in the carbon cycle variables more difficult. The CMIP6 and CMIP6 ensemble has been assessed compared to observations by \citet{Flato2013} and \citet{ipcc6ch3} respectively. A detailed analysis was also done by \citet{Bock2020} and references therein, who compare the surface temperature, pressure, precipitation, radiation, and clouds of CMIP3, CMIP5 and CMIP6 historical simulations for annual means. The CMIP6 models show better correlations for these variables than the CMIP5 models for all parameters, with smaller improvements for variables such as temperature which were already well represented in previous CMIP phases. However, the model spread is not significantly reduced but instead largely remains the same. Here, we expand the analysis to the carbon cycle. However it should be mentioned, that many carbon cycle processes are affected by physical variables on much smaller timescales, such as timing of precipitation throughout the day, or if surface temperatures fall below the freezing point at any time of the day may impact the growth of plants more than their monthly means suggest. This study uses monthly mean data which does not resolve many of these important events and thus does not investigate the impact of physical variables further, as any analysis would still 
be lacking many possible impacts of sub-frequency effects. Future studies using higher frequency data will put more emphasis also on the physical drivers of carbon processes.

\subsection{Leaf Area Index}
The Leaf Area Index (LAI) is the ratio of one-sided leaf area per unit ground area \citep{Anav2013} as a measure of the canopy structure. Models use LAI to calculate the photosynthetic uptake of the total canopy, also known as gross primary productivity (GPP). While LAI is an important building block for the carbon cycle, it was also one of the weaknesses of the carbon cycle in the CMIP5 ensemble and tended to be overestimated \citep{Anav2013, Anav2013_LAI}.

\begin{figure*}[t]
\includegraphics[width=18cm]{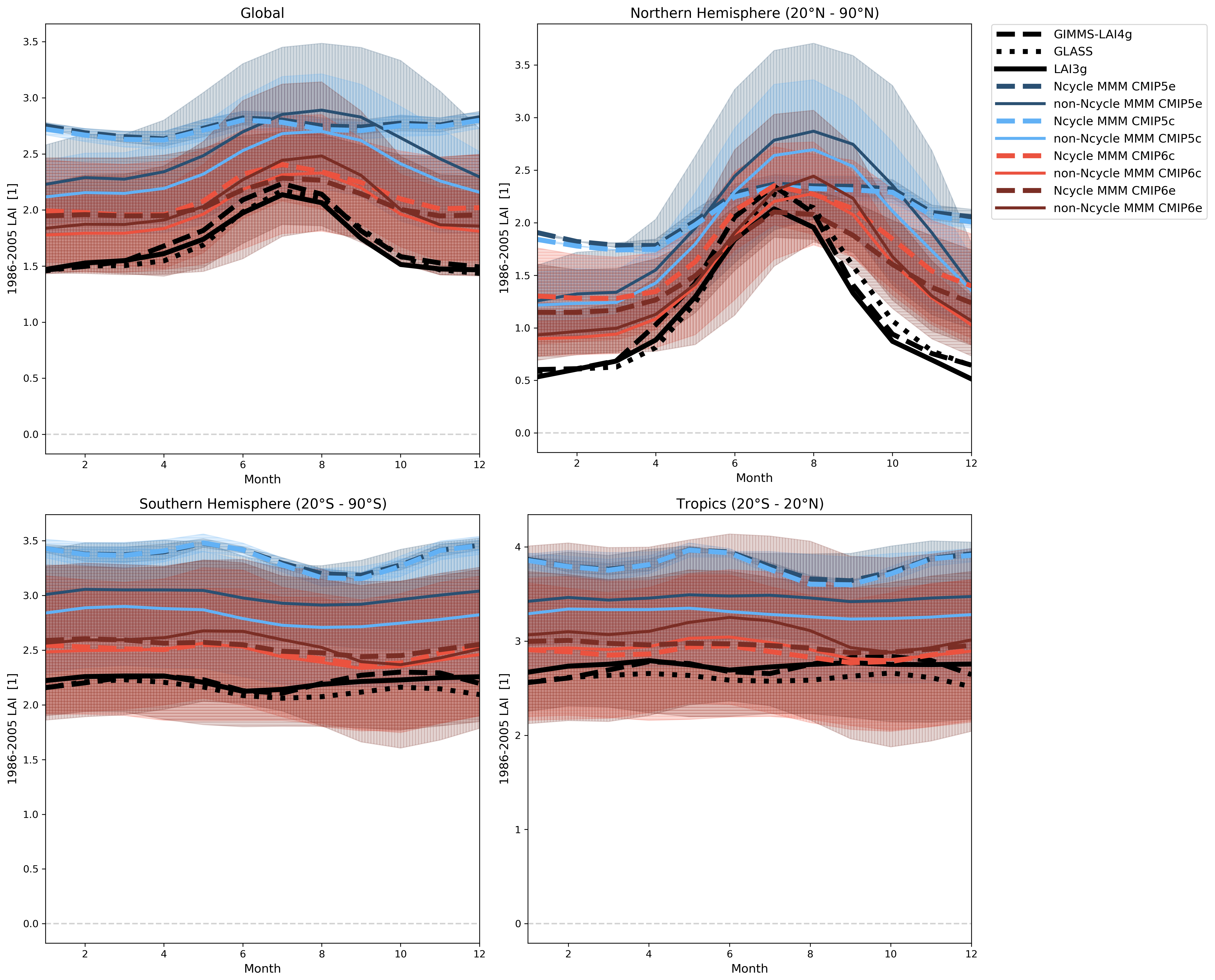} %originally width=12cm
\caption{Seasonal Cycle of leaf area index (LAI) for a climatological mean of 1986-2005 for different regional averages: Global, Northern Hemisphere (20$^\circ$N - 90$^\circ$N), Southern Hemisphere (20$^\circ$S - 90$^\circ$S) and Tropics (20$^\circ$S - 20$^\circ$N). The reference data sets (LAI3g, solid line; LAI4g, dashed line; GLASS, dotted line) are shown in black, while the MMMs for the different project-experiment combinations are denoted by blue for CMIP5, red for CMIP6, with darker colors for the emission driven simulations (dark blue CMIP5e, dark red CMIP6e) and lighter colors for concentration driven simulations (light blue CMIP5c, light red CMIP6c). MMMs derived from models with coupled nitrogen cycle (Ncycle) are dashed, while solid lines represent MMMs of models without coupled nitrogen cycle (non-Ncycle). The shading represents the standard deviation of the MMMs, with vertical hatching for models without and horizontal hatching for models with coupled nitrogen cycle. For comparison with the reference data which contain many missing values, a common mask was applied to all data sets, removing values where any data set is missing a value.}
\label{fig:lai_cycle_allproj}
\end{figure*}

Carbon uptake by land follows a pronounced seasonal cycle, with CO$_{2}$ removed from the atmosphere through plant photosynthesis and released back through plant and soil respiration. With LAI describing the canopy structure and more plants thriving in summer, it is strongly linked to the seasonal cycle of atmospheric CO$_{2}$. The seasonal cycles for LAI for CMIP5 and CMIP6 MMMs for concentration and emission driven simulations are shown in Figure \ref{fig:lai_cycle_allproj}, split into models with (Ncycle) and without interactive nitrogen cycle (non-Ncycle), as well as different regional means: Global, Northern Hemisphere (20$^\circ$N - 90$^\circ$N; NH), Southern Hemisphere (20$^\circ$S - 90$^\circ$S; SH) and Tropics (20$^\circ$S - 20$^\circ$N). From here on, concentration driven simulations will be denoted by c, such as CMIP5c and CMIP6c, and emission driven simulations by e (CMIP5e, CMIP6e). The BNU-ESM and MRI-ESM1 CMIP5 models were removed from the MMM due to featuring an unrealistically high mean LAI, almost doubling the LAI of the reference data and the other models in the SH and the tropics. A common mask is applied to all data sets which includes all missing values in any data set to allow for direct comparison between the models and reference data sets. This increases the LAI compared to unmasked regional means, as missing values are more common in desert and mountainous regions with low LAI (Fig. \ref{fig:lai_maps_ref}). 
The chosen reference data sets LAI3g, LAI4g, and GLASS agree well across all regions, which was to be expected as they are all based on the same raw satellite data from AVHRR and MODIS. \citet{Xiao2017} found that the GLASS product outperformed other products, which included LAI3g, when compared to LAI from high-resolution reference maps. As such, GLASS, which is also the reference data set with the largest coverage, will be considered the main reference data set for our analysis of LAI.

There is a strong seasonal cycle in the NH, dominated by high-latitude vegetation in Eurasia and North America. The NH seasonal cycle is the dominant contribution to the global mean due to the higher relative land fraction in the higher latitudes of the global north compared to the global south. The tropics do not show a seasonal cycle due to the absence of strong seasons, while the vegetation in the SH is dominated by forests closer to the tropics, which also lack strong seasonality. All models overestimate LAI in all regions, but the CMIP6 models reproduce the reference data better than the CMIP5 models. According to \citet{Anav2013_LAI}, the overestimation in the mid-latitudes is likely partly due to a wet bias and its control on soil moisture, a saturation of satellite instrumentation, and missing parametrizations of disturbances. The annual mean precipitation wet bias is minimally reduced in CMIP6 \citep{Bock2020} and new parametrizations such as nutrient limitations through interactive nitrogen cycle have been introduced in some models, leading to a reduced LAI in the CMIP6 MMMs compared to CMIP5. The seasonal cycle in the NH is reproduced, but while the CMIP5 non-Ncycle models reproduce the amplitude well with a positive offset of approximately 0.7\,m$^2$/m$^2$, the CMIP6 non-Ncycle models are better at reproducing the peak value. Both CMIP5 Ncycle models (CESM1-BGC and NorESM1-ME) use the CLM4 land model with known issues regarding LAI, such as underestimating LAI in dry regions due to elevated CO$_2$ and overestimating LAI in moist regions \citep{Lee2013}, as well as an unrealistically strong nitrogen limitation \citep{Wieder2019}, hindering plant growth. This leads to these models showing a larger overestimation in LAI in both the Southern Hemisphere and Tropics dominated by moist rainforests. Additionally, the seasonal cycle amplitude is strongly reduced in the NH while the mean LAI is larger than for the reference data sets. Both CMIP projects show a weakness in simulating the end of the growing season, shown by the later decline of LAI in winter, which also leads to a smaller seasonal cycle amplitude, consistent with the findings of \citet{Park2021}. The drawdown in autumn signifying the end of the growing season is smaller in Ncycle models compared to non-Ncyle models. The differences between the concentration and emission driven simulations are small, with models participating in both simulations having very similar results (individual models not shown). Larger differences occurring here and in later analysis between the concentration and emission driven simulations are likely due to the different subset of models in the historical simulations, and not due to the experiment design.

\begin{figure*}[t]
\includegraphics[width=18cm]{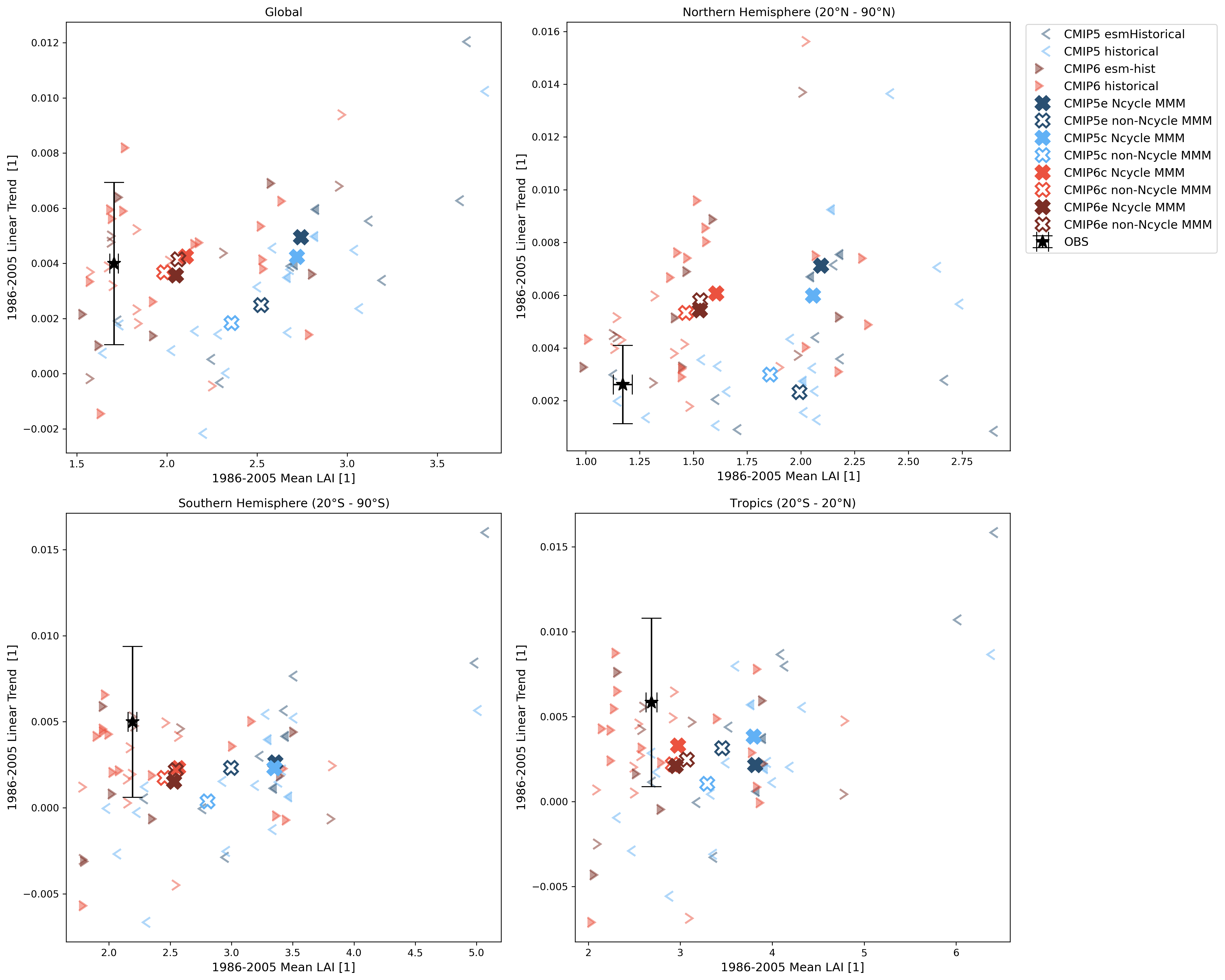} %originally width=12cm
\caption{Mean and Trend of LAI computed over 1986-2005 for different regions: Global, Northern Hemisphere (20$^\circ$N - 90$^\circ$N), Southern Hemisphere (20$^\circ$S - 90$^\circ$S) and Tropics (20$^\circ$S - 20$^\circ$N). The mean of the reference data sets (LAI3g, LAI4g and GLASS) is denoted by a black star, with errorbars for the standard deviation. Models for project-experiment combinations are shown with a single symbol each, blue smaller than (<) sign for CMIP5 and red greater than (>) sign for CMIP6, with darker colors for emission driven and lighter colors for concentration driven simulations, as well as Ncycle models being denoted with a filled symbol. MMMs are depicted with cross symbols, filled for Ncycle MMMs and using the color assigned to their project-experiment combination.}
\label{fig:lai_scatter}
\end{figure*}

\begin{figure*}[t]
\includegraphics[width=18cm]{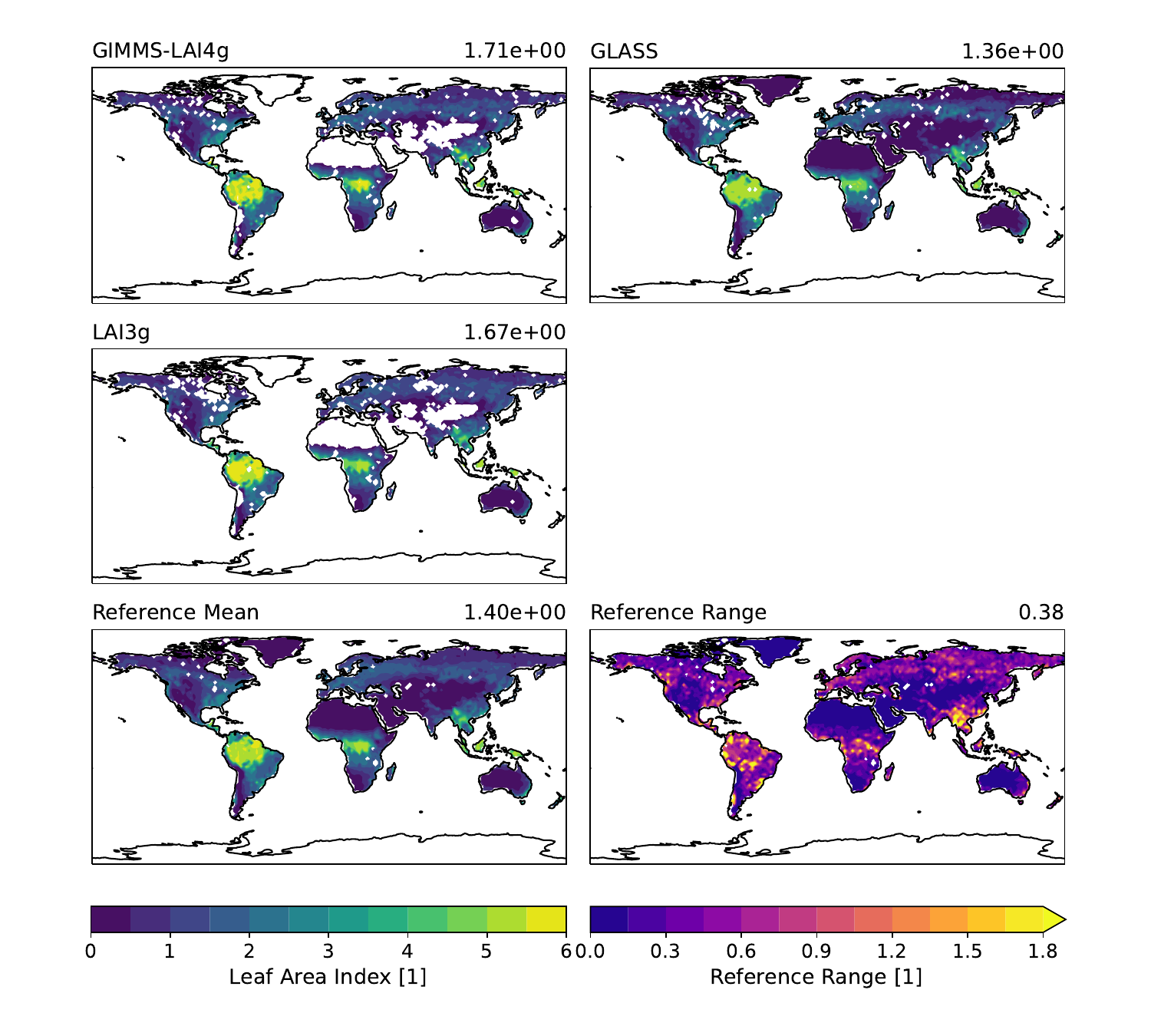} %originally width=12cm
\caption{Global maps of LAI averaged over 1986-2005 for all reference data sets, as well as the mean and range between lowest and highest values per grid cell of the reference data sets.}
\label{fig:lai_maps_ref}
\end{figure*}

Figure \ref{fig:lai_scatter} shows the mean and trend of LAI averaged over 1986-2005 and depicts project-experiment simulations with one type of marker each for a better overview. Filled symbols denote models with nitrogen cycle while the mean of the reference data is shown with black markers, with error bars showing the standard deviation of the reference data. Exact numbers for all data sets are found in the supplementary information (Tables S1-S2). The errors given refer to the standard deviation of the mean as a measure for the interannual variability (IAV), while the standard error of the trend is the error of the linear regression calculating the trend. For the individual reference data, LAI3g and LAI4g agree well in mean and trend, while GLASS agrees with their mean but shows a significantly higher trend in all regions, leading to the large trend error bar. \citet{Xiao2017} analyzed the trend of several LAI products for different biome types for 1982-2011 and found GLASS to have significantly higher trends in savannahs and shrubs compared to LAI3g, but lower trends in deciduous broadleaf forests, evergreen needleleaf forests and even a negative trend for deciduous needleleaf forests, while grasses, cereal crops and evergreen broadleaf forests trends are similar for GLASS and LAI3g. This results in larger trend differences in the SH (0.009\,m$^2$/m$^2$ yr$^{-1}$) and Tropics (0.010\,m$^2$/m$^2$ yr$^{-1}$) than the NH (0.003\,m$^2$/m$^2$ yr$^{-1}$). MMMs span the range of 1.98 $\pm$ 0.02\,m$^2$/m$^2$ (non-Ncycle CMIP6c) to 2.74 $\pm$ 0.05\,m$^2$/m$^2$ (Ncycle CMIP5e), showing a significant overestimation compared to the reference data, with the overestimation of the CMIP5 MMMs of 0.7\,m$^2$/m$^2$ reduced by half for CMIP6 MMMs, as seen in Figure \ref{fig:lai_cycle_allproj}. This large improvement for the CMIP6 models is not related to the difference between Ncycle and non-Ncycle models as their LAI MMMs means are comparable. Due to the large difference in trends in the reference data sets, the global mean trends of all CMIP MMMs lie within the range of the reference data. The model trends range between slightly negative -0.0022 $\pm$ 0.0010 \,m$^2$/m$^2$ yr$^{-1}$ (IPSL-CM5A-LR CMIP5c) to strong positive 0.0120 $\pm$ 0.0013 \,m$^2$/m$^2$ yr$^{-1}$ (BNU-ESM CMIP5e). As BNU-ESM was not considered in the MMM due to its large mean LAI compared to all the other models, the highest global LAI trend considered for the MMM is CMIP6c MRI-ESM2-0 at 0.0094 $\pm$ 0.0019 \,m$^2$/m$^2$ yr$^{-1}$, which is significantly larger than the trend in its CMIP6e simulation at 0.0068 $\pm$ 0.0019 \,m$^2$/m$^2$ yr$^{-1}$. The MMMs range between 0.0018 $\pm$ 0.0004 \,m$^2$/m$^2$ yr$^{-1}$ (non-Ncycle CMIP5c) and 0.0050 $\pm$ 0.0015 \,m$^2$/m$^2$ yr$^{-1}$ (Ncycle CMIP5e). Unlike the mean, the LAI trend does not show a strong difference between CMIP5 projects, nor any other grouping method we employed. The CMIP6 models only show a slightly smaller range in trend compared to the CMIP5 models, but more CMIP5 models have a lower trend compared to the reference data sets than CMIP6 models.

The other regions reflect these overall MMM comparisons as well. CMIP6 means are closer to the reference data than CMIP5 in all regions, albeit still overestimating LAI, and agree well with each other no matter the experiment or Ncycle status. The only exception is the NH CMIP6c Ncycle MMM, which shows a larger mean at 1.71 $\pm$ 0.04\,m$^2$/m$^2$ than the other three CMIP6 groupings with means between 1.47$\pm$ 0.03\,m$^2$/m$^2$ and 1.53 $\pm$ 0.04\,m$^2$/m$^2$. This is due to the CMCC-ESM2 and CMCC-CM2-SR5 models, which show a much higher LAI in the NH compared to the reference data, but fit well in the other regions. These two models use the CLM4.5 land model, which \citet{Li2022} found to have a far longer peak growing season and to overestimate LAI in boreal forests compared to MODIS reference data, consistent with our results. The CLM5 models (CESM2, CESM2-WACCM, NorESM2-LM, NorESM2-MM) perform much better in mean LAI than CLM4.5 in the NH, but are still overestimating LAI compared to the reference data. The trend in the NH for the reference data set are 0.0014 $\pm$ 0.0003\,m$^2$/m$^2$ yr$^{-1}$ (LAI4g), 0.0018 $\pm$ 0.0008\,m$^2$/m$^2$ yr$^{-1}$ (LAI3g) and 0.0047 $\pm$ 0.0013 \,m$^2$/m$^2$ yr$^{-1}$ (GLASS), with the models showing a much larger range from 0.0008 $\pm$ 0.0008\,m$^2$/m$^2$ yr$^{-1}$ (GFDL-ESM2M CMIP5e), with a significantly larger trend in CMIP5c at 0.0057 $\pm$ 0.0006\,m$^2$/m$^2$ yr$^{-1}$, to 0.0156 $\pm$ 0.0023\,m$^2$/m$^2$ yr$^{-1}$ (MRI-ESM2-0 CMIP6c), although most models have a trend below 0.01\,m$^2$/m$^2$ yr$^{-1}$. For the MMMs, only the CMIP5 non-Ncycle MMMs fall between the reference data with 0.0023 $\pm$ 0.0003\,m$^2$/m$^2$ yr$^{-1}$ for non-Ncycle CMIP6c and 0.0030 $\pm$ 0.0003\,m$^2$/m$^2$ yr$^{-1}$ for non-Ncycle CMIP6e. The other MMMs show a larger trend than the reference data but comparable to each other, ranging from 0.0053 $\pm$ 0.0004\,m$^2$/m$^2$ yr$^{-1}$ (non-Ncycle CMIP6c) to 0.0061 $\pm$ 0.0004\,m$^2$/m$^2$ yr$^{-1}$ (Ncycle CMIP6c), with a slightly larger value of 0.0071 $\pm$ 0.0012\,m$^2$/m$^2$ yr$^{-1}$ for Ncycle CMIP5e. CMIP6 MMM LAI mean ($\approx$ 2.5\,m$^2$/m$^2$ SH, 2.9\,m$^2$/m$^2$ tropics) and trend ($\approx$ 0.002\,m$^2$/m$^2$ yr$^{-1}$ SH, 0.002\,m$^2$/m$^2$ yr$^{-1}$ tropics) agree well with the LAI3g and LAI4g (mean 2.2\,m$^2$/m$^2$ SH, 2.7\,m$^2$/m$^2$ tropics, trend 0.002\,m$^2$/m$^2$ yr$^{-1}$ SH, 0.002\,m$^2$/m$^2$ yr$^{-1}$ tropics) reference data in the tropics and the SH, while the CMIP5 MMMs overestimate the mean by $\approx$ 0.7\,m$^2$/m$^2$ in the SH and $\approx$ 0.6\,m$^2$/m$^2$ in the tropics for non-Ncycle, as well as $\approx$ 1.2\,m$^2$/m$^2$ in the SH and $\approx$ 1.1\,m$^2$/m$^2$ in the tropics for Ncycle MMMs, but show a similar trend to LAI3g, LAI4g, and CMIP6 MMMs. The larger mean LAI in the CLM4 (CMIP5 Ncycle MMMs) can be traced back to the overestimation of LAI in moist regions mentioned before. Some models show a significant negative trend in LAI in the SH and the tropics resulting in a globally negative trend even with a positive trend in the NH.

\begin{figure*}[t]
\includegraphics[width=12cm]{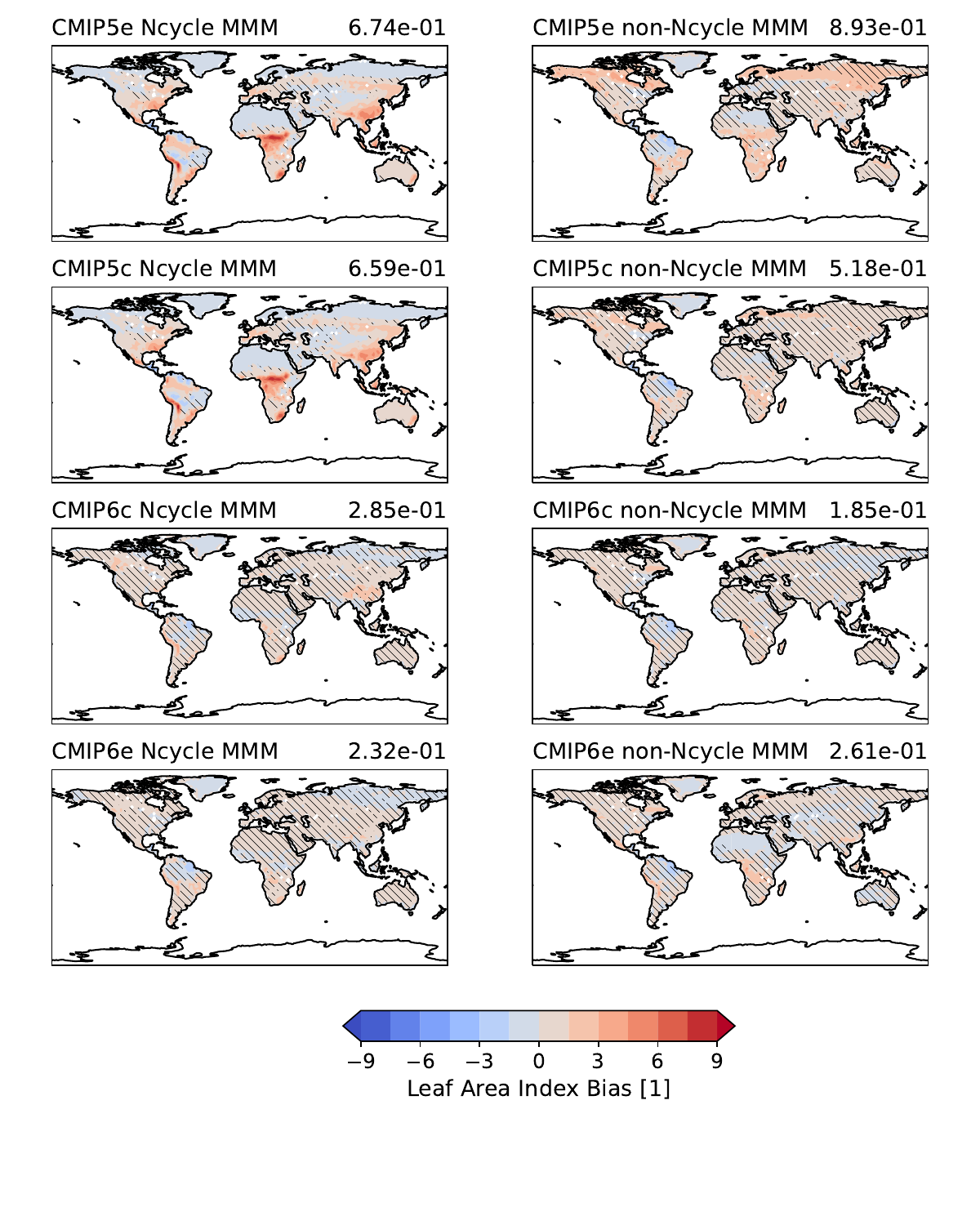} %originally width=12cm
\caption{Global maps of LAI bias for 1986-2005 with respect to the reference data set mean shown in Figure \ref{fig:lai_maps_ref}. The panels show the MMMs of the models with (left) and without (right) coupled nitrogen cycle for the different project-experiment combinations. The hatching represents the areas where the MMM of the models and reference mean agree within the MMM std.}
\label{fig:lai_maps_means}
\end{figure*}

Maps of the LAI reference data are found in Figure \ref{fig:lai_maps_ref}, without the common mask to see the different coverages. Coverage of the different reference data sets varies a lot due to different quality control criteria and algorithms, with most missing values found in desert or mountainous regions such as the Sahara and the Himalayas. Additionally, a mean of the reference data and the range of the reference data per grid cell is shown, along with the global mean of the values in the upper right corner. GLASS has a larger coverage over desert and mountainous regions, which are regions with low plant coverage and thus low LAI, resulting in a lower global mean LAI of 1.36\,m$^2$/m$^2$ compared to 1.67\,m$^2$/m$^2$ and 1.71\,m$^2$/m$^2$ from LAI3g and LAI4g respectively. This underlines the importance of the common mask used for figures \ref{fig:lai_cycle_allproj} and \ref{fig:lai_scatter} to obtain comparable results. The different data sets show the same pattern of LAI distribution, with the absolute values ranging between 1 and 6\,m$^2$/m$^2$, while the differences are below 2\,m$^2$/m$^2$ with the largest difference occurring in tropical rain forests and northern high latitudes, the regions with the largest absolute LAI values. For a gridcell bias comparison of the different model groupings (Fig. \ref{fig:lai_maps_means}) a combined reference data set was computed as the mean of the other reference data sets. Due to the different coverages, some areas are only calculated from the GLASS data, while others are an average of all three data sets. The range of values per gridcell going into the combined data set is plotted in the lower right of Figure \ref{fig:lai_maps_ref}. The largest difference occur in the areas with larger LAI as the tropical rainforests followed by boreal forests, with a global mean average range of 0.38\,m$^2$/m$^2$. For the MMM bias maps shown in Figure \ref{fig:lai_maps_means} hatching is added where the MMMs agree with the reference mean within the MMM standard deviation. CMIP5 Ncycle MMMs show the issue of CLM4 in overestimating LAI in wet regions, with LAI in tropical rainforests almost doubling the reference value, while drier regions show a significant negative bias. While this results in a global mean bias of 0.66 to 0.67\,m$^2$/m$^2$ smaller than 0.89\,m$^2$/m$^2$ for CMIP5e non-Ncycle MMMs, it is still the worst performing model grouping when considering a gridcell basis due to its extreme biases in both directions. The hatching showing the agreement can be ignored in this case, as only two models contributed to the MMM std. The CMIP5e non-Ncycle MMM shows a strong overestimation across the northern latitudes besides Greenland. This can be attributed to the GFDL-ESM2G and GFDL-ESM2M models, which are known to have established coniferous trees in areas which should contain tundra or cold deciduous trees, as its vegetation spin-up only coniferous trees are allowed to grow in cold regions, but not grasses or deciduous trees which would have a lower LAI \citep{Anav2013_LAI}. While the GFDL models show this problem in both CMIP5c and CMIP5e, due to the larger number of MMMs contributing to the concentration driven simulations, their effect is reduced. In CMIP6, GFDL-ESM4 still has a large positive LAI bias throughout these areas, but it is significantly reduced compared to its CMIP5 predecessor. The second prominent overestimation is around the tropical rainforests, where models like BNU-ESM, MRI-ESM1, and in lesser extent also the GFDL models extend the LAI hotspot to larger areas around it compared to the reference mean. In CMIP5c HadGEM2-CC and HadGEM2-ES also show this overestimation. CMIP5c non-Ncycle MMM shows no special bias patterns, but instead a general overestimation in almost all areas with hatching present throughout the globe. CMIP6 MMMs show a reduced mean bias of less than half the CMIP5 overestimation, with almost no pronounced patterns and a bias reduction in all areas, with the largest improvements found in the northern high latitudes. The largest bias is in southeast asia for CMIP6c Ncycle MMMS which can be tracked to the CMCC-ESM2 and CMCC-CM2-SR5 models, and makes the mean bias of the CMIP6c Ncycle MMM higher than that of CMIP6c non-Ncycle MMM. Otherwise, the bias pattern looks similar for CMIP6 Ncycle and non-Ncycle models.

While CMIP6 LAI has improved compared to CMIP5, especially a significantly reduced mean bias, a general overestimation of LAI remains, along with issues of correctly reproducing the length of the growing season in the northern hemisphere, and a large model spread in mean and trend LAI. Neither the introduction of an interactive nitrogen cycle nor the comparison between emission and concentration driven simulations show large differences in the overall quality of the CMIP6 simulations for LAI.

\subsection{Gross Primary Productivity}

\begin{figure*}[t]
\includegraphics[width=18cm]{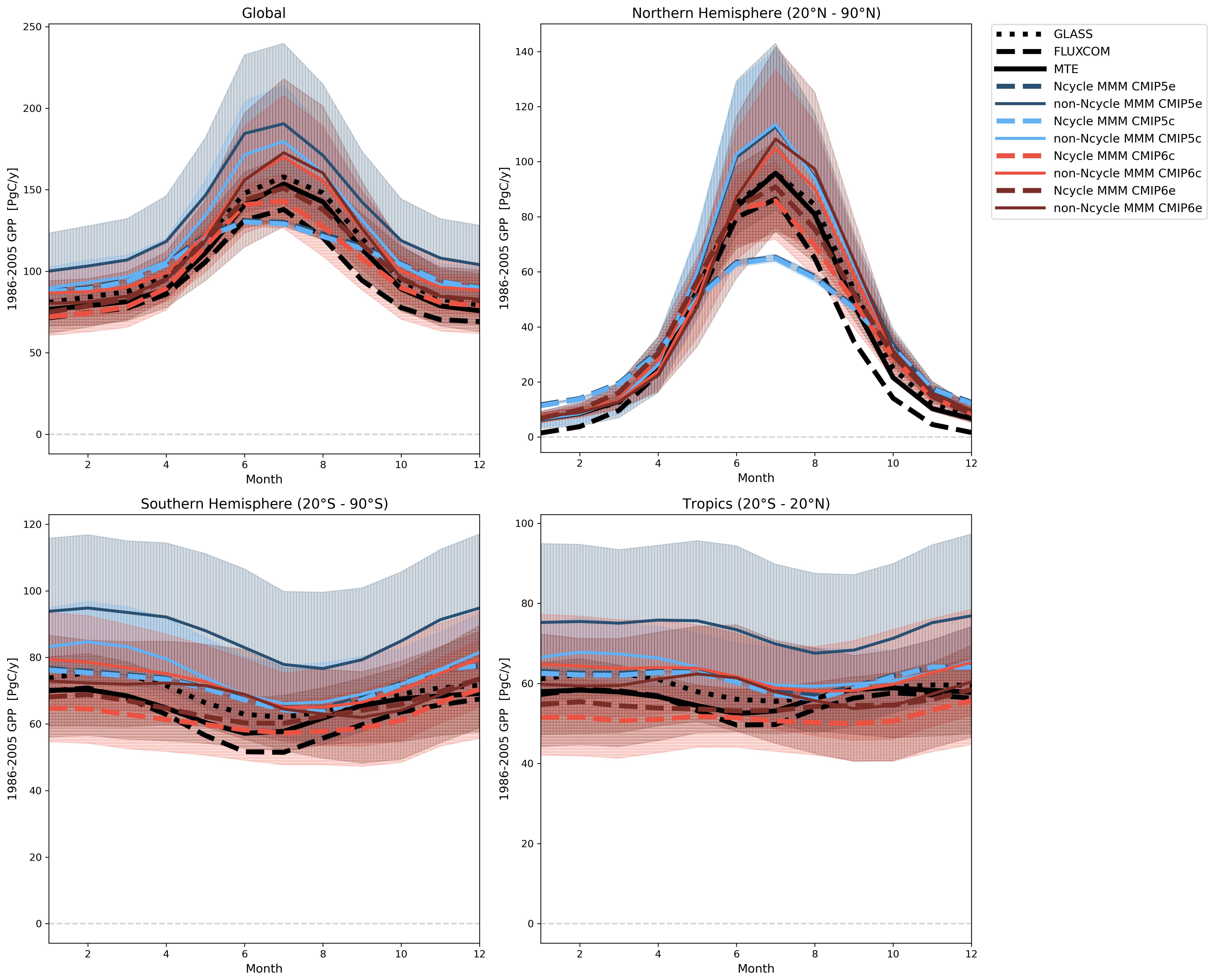} %originally width=12cm
\caption{As Figure \ref{fig:lai_cycle_allproj} but for gross primary production using GLASS, FLUXCOM and MTE reference data. Additionally, the regional GPP is calculated as the area weighted sum instead of the mean of the gridcells used for LAI.}
\label{fig:gpp_cycle_allproj}
\end{figure*}

Gross primary productivity (GPP) represents the CO$_2$ uptake on land due to photosynthesis. This was one of the biggest weaknesses of the CMIP5 ensemble, with most models overestimating photosynthesis as well as leaf area index \citep{Anav2013}. The seasonal cycle of GPP (Fig. \ref{fig:gpp_cycle_allproj}) shows good agreement between the GLASS and MTE reference data, while the FLUXCOM data shows a lower GPP in all regions, as well as a shorter growing season in the Northern Hemisphere. All models reproduce a similar shape of the NH seasonal cycle to the GLASS and MTE data in both model generations and experiments. As found in \citet{Anav2013} the CMIP5 non-Ncycle models overestimate GPP in all regions, while the CMIP5 Ncycle models strongly underestimate the peak of the seasonal cycle in the NH. CMIP6 models perform better than CMIP5, but while the CMIP6 non-Ncycle models still overestimate the GPP peak in summer similarly to CMIP5 for the NH, the Ncycle models show a very good agreement for both CMIP6c and CMIP6e. Nitrogen limitation is stronger in northern latitudes through boreal forests and tundra \citep{du2020}, as compared to tropical and subtropical forests, which are more limited by phosphorus. However, from the CMIP models used in this study, only ACCESS-ESM1-5 includes an interactive phosphorus cycle. It therefore makes sense that Ncycle models show a decreased GPP compared to non-Ncycle models in the NH, closer to reference data. The model spread remains large in CMIP6, albeit smaller for Ncycle models, which is denoted by horizontal hatching. Following LAI, there is no strong discernible seasonal cycle in neither the SH nor the Tropics, but the CMIP6 models are closer to the mean GPP than the CMIP5 models, with lower values for Ncycle models.

\begin{figure*}[t]
\includegraphics[width=18cm]{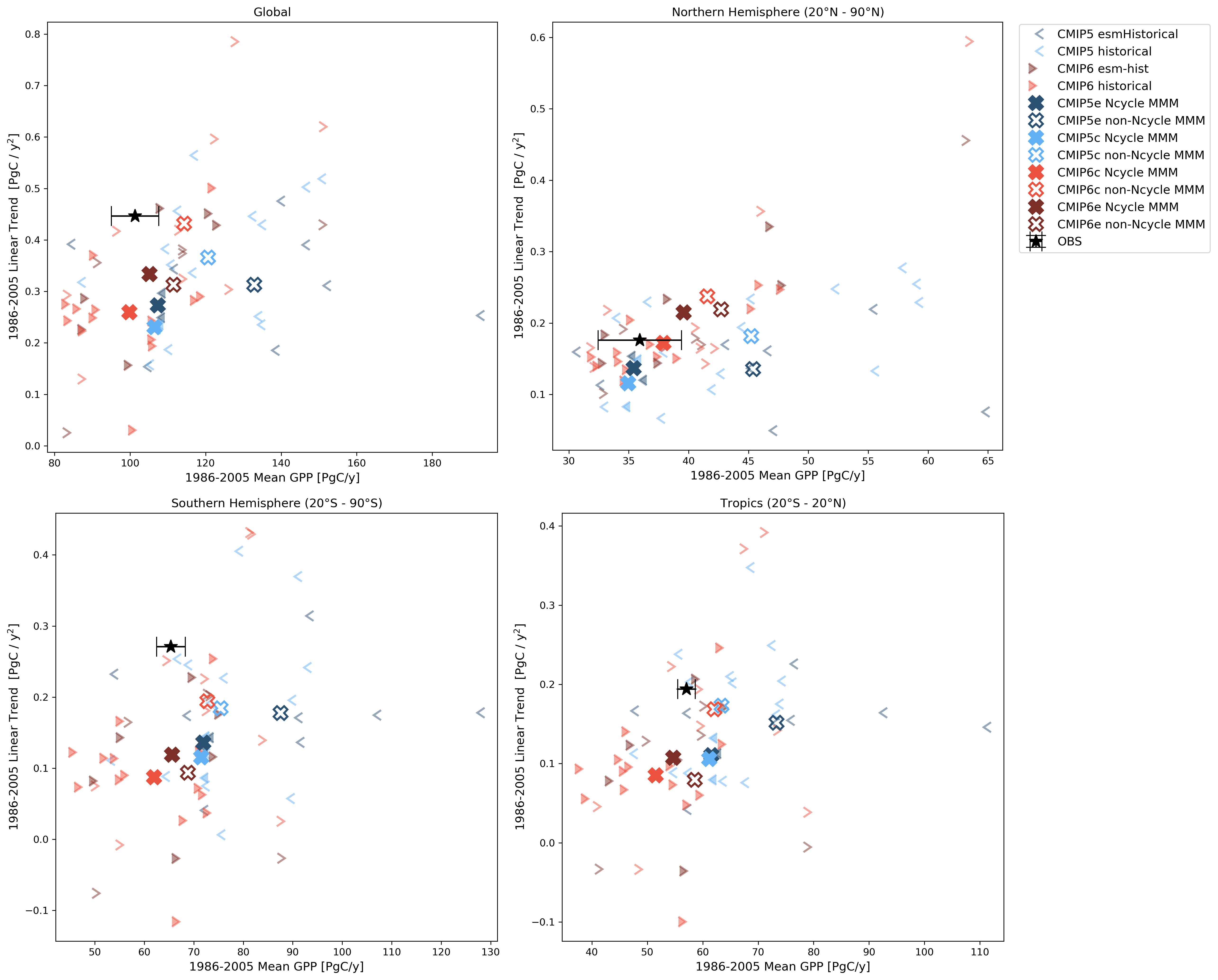} %originally width=12cm
\caption{As Figure \ref{fig:lai_scatter} but for gross primary production using GLASS, FLUXCOM and MTE reference data.}
\label{fig:gpp_scatter}
\end{figure*}

The temporal mean and linear trend of the spatial sums for GPP during the time period 1986-2005 is shown in Figure \ref{fig:gpp_scatter}. MTE and FLUXCOM data is known to have an unrealistic non-existent trend (0.01\,PgC yr$^{-2}$ globally) due to the assumption of an unchanging average CO$_2$ level \citep{Anav2015} in these data sets. As such, the model trend should not be compared to the trend of these two reference data sets, and we have omitted these data sets from the calculation of the reference trend. The mean GPP of all regions of these data sets along with all other numerical values from the plot can be found in Tables S3-S4. The trend of GLASS (0.45 $\pm$ 0.09\,PgC yr$^{-2}$ globally) is closely linked to the high trend of LAI GLASS, one of the main influences on GPP. The reference data sets agree well in mean GPP with the largest difference being a lower mean for FLUXCOM in the NH. Globally the reference values range from 93.0 $\pm$ 0.4\,PgC yr$^{-1}$ for FLUXCOM and 102.6 $\pm$ 1.2\,PgC yr$^{-1}$ for MTE to 108.3 $\pm$ 3.4\,PgC yr$^{-1}$ for GLASS. The models show a large range from 83.5 $\pm$ 2.5\,PgC yr$^{-1}$ (BNU-ESM CMIP5e) to 152.0 $\pm$ 4.6\,PgC yr$^{-1}$ with an even larger mean for MRI-ESM1 CMIP5e as an outlier, which shows a large GPP in all regions. The CMIP5 models are on the higher side of this range, as seen by the MMMs of 120.6 $\pm$ 2.2\,PgC yr$^{-1}$ (non-Ncycle CMIP5c) and 132.9 $\pm$ 2.1\,PgC yr$^{-1}$ (non-Ncycle CMIP5e) with the Ncycle CMIP5 models much lower at 106.5 $\pm$ 1.6\,PgC yr$^{-1}$ (Ncycle CMIP5c) and 107.3 $\pm$ 1.9\,PgC yr$^{-1}$ (Ncycle CMIP5e) due to their underestimation in the NH as seen in Figure \ref{fig:gpp_cycle_allproj}. The CMIP6 Ncycle MMMs agree very well with the reference data, while the CMIP6 non-Ncycle MMMs show a larger mean GPP. The global trend of GLASS (0.45 $\pm$ 0.09\,PgC yr$^{-2}$) is positive, which is well matched by the non-Ncycle CMIP6c MMM with the other MMMs showing a smaller trend. In the NH, more CMIP5 models overestimate the mean GPP than CMIP6 models. The MMMs span a large spread, with the Ncycle MMMs showing a lower mean GPP than the non-Ncycle MMMs which are overestimating GPP compared to the reference data. The models are clustered around the GLASS trend, with outliers for MRI-ESM2-0 in both its CMIP6c and CMIP6e runs. The CMIP6 non-Ncycle MMM shows a higher trend than the other MMMs due to the MRI-ESM2-0 outliers. In the SH mean GPP, the CMIP6 MMMs match the reference data well with lower values for Ncycle than for non-Ncycle MMMs. The CMIP5 MMMs are slightly above these, with the non-Ncycle CMIP5e having a much larger value due to the outlier of MRI-ESM1 and FIO-ESM both with values well above 100\,PgC yr$^{-1}$. Compared to the GLASS trend of 0.27 $\pm$ 0.07\,PgC yr$^{-2}$, the MMMs underestimate the trend. The distribution in the tropics is very similar as the SH. The GLASS trend of 0.19 $\pm$ 0.05\,PgC yr$^{-2}$ is underestimated by MMMs. In summary, the Ncycle MMM shows a better performance than the non-Ncycle MMM in the NH, while it shows a slight underestimation in the tropics and a similar performance in the SH. The model spread over the trend in GPP stays similar throughout the model generations, with the mean trend being largely consistent with the GLASS reference data set in all regions but underestimated everywhere but in the NH.

\begin{figure*}[t]
\includegraphics[width=18cm]{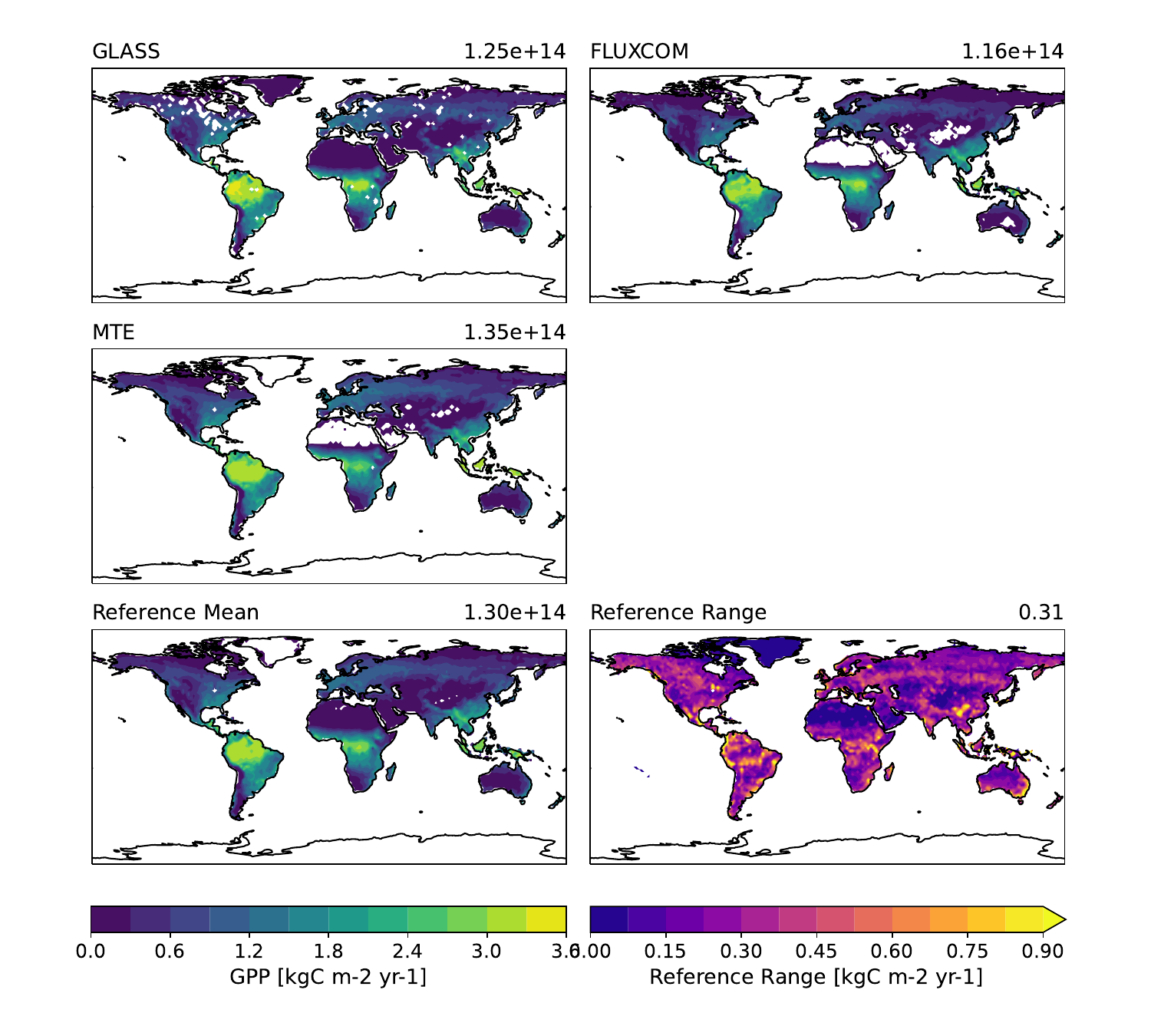} %originally width=12cm
\caption{Global maps of GPP averaged over 1986-2005 for all reference data sets (GLASS, FLUXCOM and MTE), as well as the mean and range between lowest and highest values per grid cell of the reference data sets. The number in the top right denotes the global GPP flux.}
\label{fig:gpp_mapsref}
\end{figure*}

\begin{figure*}[t]
\includegraphics[width=12cm]{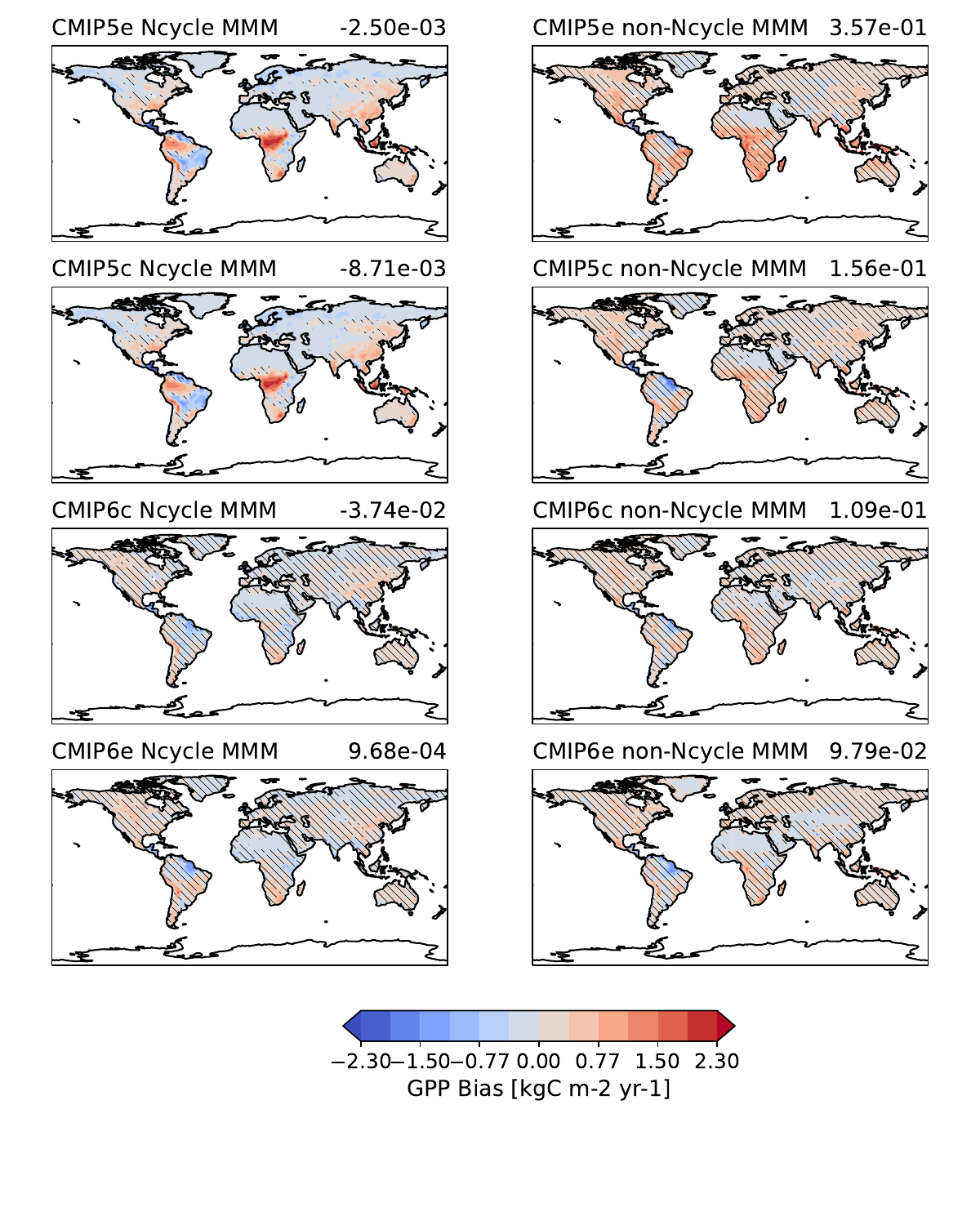} %originally width=12cm
\caption{Global maps of GPP bias for 1986-2005 with respect to the reference data set mean shown in Figure \ref{fig:gpp_mapsref}. The panels show the MMMs of the models with (left) and without (right) coupled nitrogen cycle for the different project-experiment combinations. The hatching represents the areas where the MMM of the models and reference mean agree within the MMM std, while the number in the top right denotes the global mean bias.}
\label{fig:gpp_mapsmmm}
\end{figure*}

\begin{figure*}[t]
\includegraphics[width=18cm]{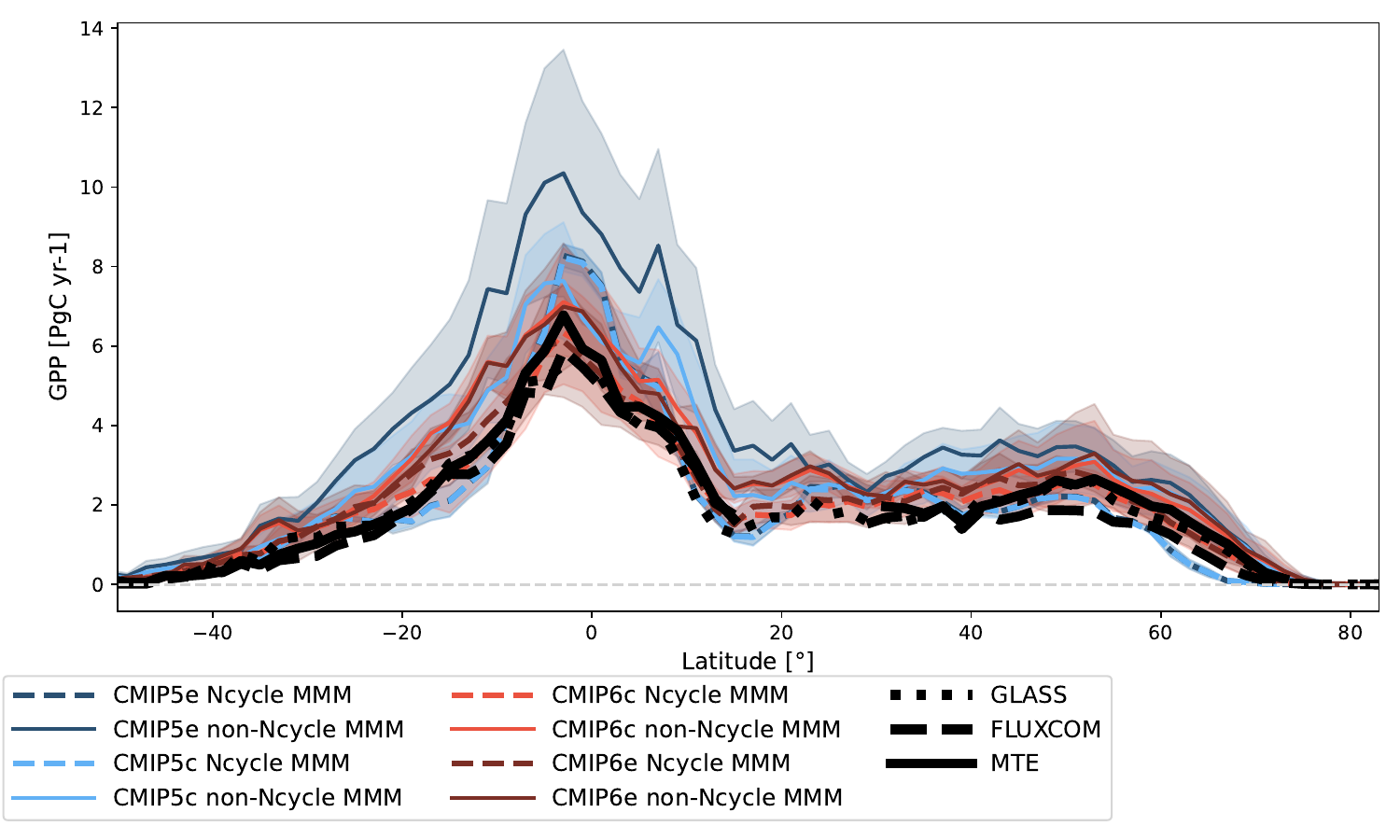} %originally width=12cm
\caption{Area weighted zonal sums of gross primary production and the reference data sets GLASS, FLUXCOM and MTE. No common masking is applied, but latitudes were set to missing if more than 15\% of the land grid cells contained missing data. The hatching depicts the MMM standard deviation, with a horizontal hatching for models with and vertical hatching for models without interactive nitrogen cycle.}
\label{fig:gpp_zmeans}
\end{figure*}

As for LAI, the coverage of the GLASS data is larger than for the other reference data, with missing values for FLUXCOM and MTE mainly found over the Sahara and the Himalayas (Fig. \ref{fig:gpp_mapsref}), GLASS shows a larger GPP in the tropical rainforests and boreal forests, explaining the larger mean GPP seen before. The reference mean data set has hardly any missing values left and the largest difference in the data sets are at places with the highest GPP, with discrepancies in the areas bordering the hotspots of the rainforests and boreal forests. Even though the GPP bias maps (Fig. \ref{fig:gpp_mapsmmm}) for Ncycle CMIP5 MMMs have a global mean bias almost two magnitudes smaller than the CMIP5 non-Ncycle MMMs, they show the same pattern of overestimation in wet regions and underestimation in dry regions found for LAI (Fig. \ref{fig:lai_maps_means}), underlining the strong influence of LAI on GPP. The CMIP5 non-Ncycle MMMs also shows similar patterns to the LAI bias maps, but the overestimation in the areas around the tropical rainforests is strongly reinforced, while the previous strong overestimation of LAI in the northern high latitudes for the non-Ncycle CMIP5e MMM is reduced. The CMIP5c non-Ncycle MMM additionally shows a strong underestimation at the northeastern coast of South America. The global mean bias for CMIP5 non-Ncycle MMMs lies at 3.6$\cdot$10$^{-13}$\,PgC m$^{-2}$ yr$^{-1}$ for CMIP5e and is reduced by half for CMIP5c. This bias is further reduced to approximately 1.0$\cdot$10$^{-13}$\,PgC m$^{-2}$ yr$^{-1}$ for CMIP6 non-Ncycle MMMs, which show similar bias patterns to CMIP5 non-Ncycle MMMs but overall reduction to the bias patterns, with a larger reduction in the savannah regions of Africa. The global mean bias for the CMIP6 Ncycle MMMs is further reduced to 0.01$\cdot$10$^{-13}$\,PgC m$^{-2}$ yr$^{-1}$ for CMIP6e and a negative bias of -0.4$\cdot$10$^{-13}$\,PgC m$^{-2}$ yr$^{-1}$ for CMIP6c. Both show a reduction in the northern hemisphere bias of the open shrublands, turning some into a negative bias, as well as southwest Africa, while the slight overestimation in North America remains, as well as the underestimation at the north eastern part of South America. This is summarized in Figure \ref{fig:gpp_zmeans}, which shows the zonal sums of the reference data and the MMMs. Unlike the seasonal cycle and scatterplots shown before, a common mask is not applied here, but instead values are masked out if a data set has more than 15\% of a latitudes land points set to missing values. The large overestimation of non-Ncycle CMIP5e can be seen which is reduced in non-Ncycle CMIP5c, with both showing a peak slightly north of the equator which is not seen in the reference data and which is due to the overestimation of the shrublands south of the Sahara. The CMIP6 non-Ncycle MMMs show a much better approximation across all latitudes, with a slight reduction of the bias in the NH, but still show a significant overestimation in the tropics. This is remedied in the CMIP6 Ncycle models, which show a very good agreement with the reference data across all latitudes, now with slight underestimations at high latitudes.

\subsection{Land-Atmosphere Flux}

The net carbon flux from the atmosphere into the land (net biome productivity, NBP) characterizes the balance between carbon uptake due to photosynthesis and carbon release by respiration, as well as other processes like fires and de- and afforestation. Positive values of NBP denote carbon uptake by land. The CMIP6 EC-Earth models (EC-Earth3-CC, EC-Earth-Veg) are excluded from the MMM for NBP because they show a very strong land source in December in seemingly random grid cells all over the globe. MIROC-ESM and MIROC-ESM-CHEM are also removed for a similar reason: they contain grid cells which seemingly randomly show large sources and sinks popping up in random months. Due to their appearance at random months instead of only in December like in EC-Earth, it does not influence regional means or climatologies as much, but can be seen in mean map plots very well. 

\begin{figure*}[t]
\includegraphics[width=11cm]{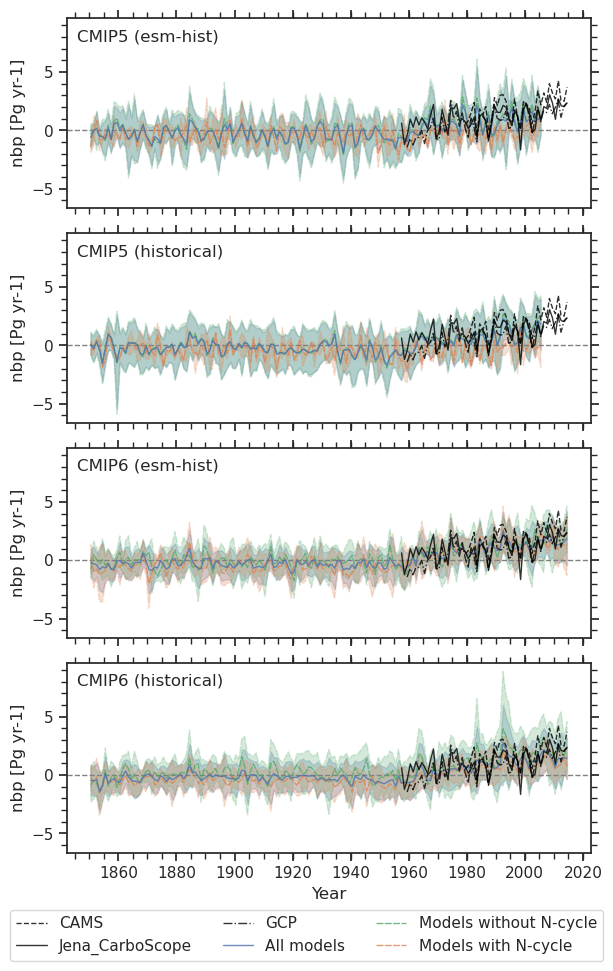} %originally width=12cm
\caption{Global Land-Atmosphere carbon flux time series for CMIP5 (top two panels) and CMIP6 (bottom two panels) concentration (panels 2 \& 4) and emission driven (panels 1 \& 3) historical simulations. Model results are separated into MMMs of models with (orange dashed line) and without interactive nitrogen cycle (green dashed line), with the MMM of all models shown in blue and the reference data sets CarboScope (solid), CAMS (dashed), and GCP (dash-dotted) shown in black. The standard deviation of the MMM is given by the shaded areas.}
\label{fig:nbp_ts}
\end{figure*}

Figure \ref{fig:nbp_ts} shows the global evolution of the land-atmosphere flux for CMIP5e, CMIP5c, CMIP6e and CMIP6c simulations in order from top to bottom. Each panel shows the MMM of all models with simulations in the respective project and experiment combinations (blue line), as well as the two MMMs of models with (dashed orange) and without (dashed green) nitrogen cycle. The colored shading represents the standard deviation of their respective MMMs. For comparison with observations, data from CAMS (dashed black), Jena CarboScope (solid black), and GCP (dash-dotted black) are added to each panel. There is a large year-to-year variability, which can also be seen in the models. All project-experiment combinations agree with all reference data sets, with the CMIP6 simulations showing a better agreement in the 1990s, during which the CMIP5 models generally underestimate the land carbon sink. From 1850 to 1970 the models do not show a significant carbon source or sink, only the CMIP6e MMM shows a small carbon source in this time period. However, the model variance as indicated by the shaded areas is large enough to be in agreement with a neutral state. Since the 1980s the land has been acting as a carbon sink which is increasing over time. This increase has previously been attributed mainly to the fertilization effect from rising atmospheric CO$_{2}$ concentrations \citep{ipcc6ch5}. In the CMIP5 simulations, Ncycle and non-Ncycle models do not show any significant differences before 1980, after which the Ncycle models show a slightly lower carbon flux. Ncycle CMIP6 models show a slightly lower land-atmosphere carbon flux over the full time period compared to non-Ncycle models.

\begin{figure*}[t]
\includegraphics[width=18cm]{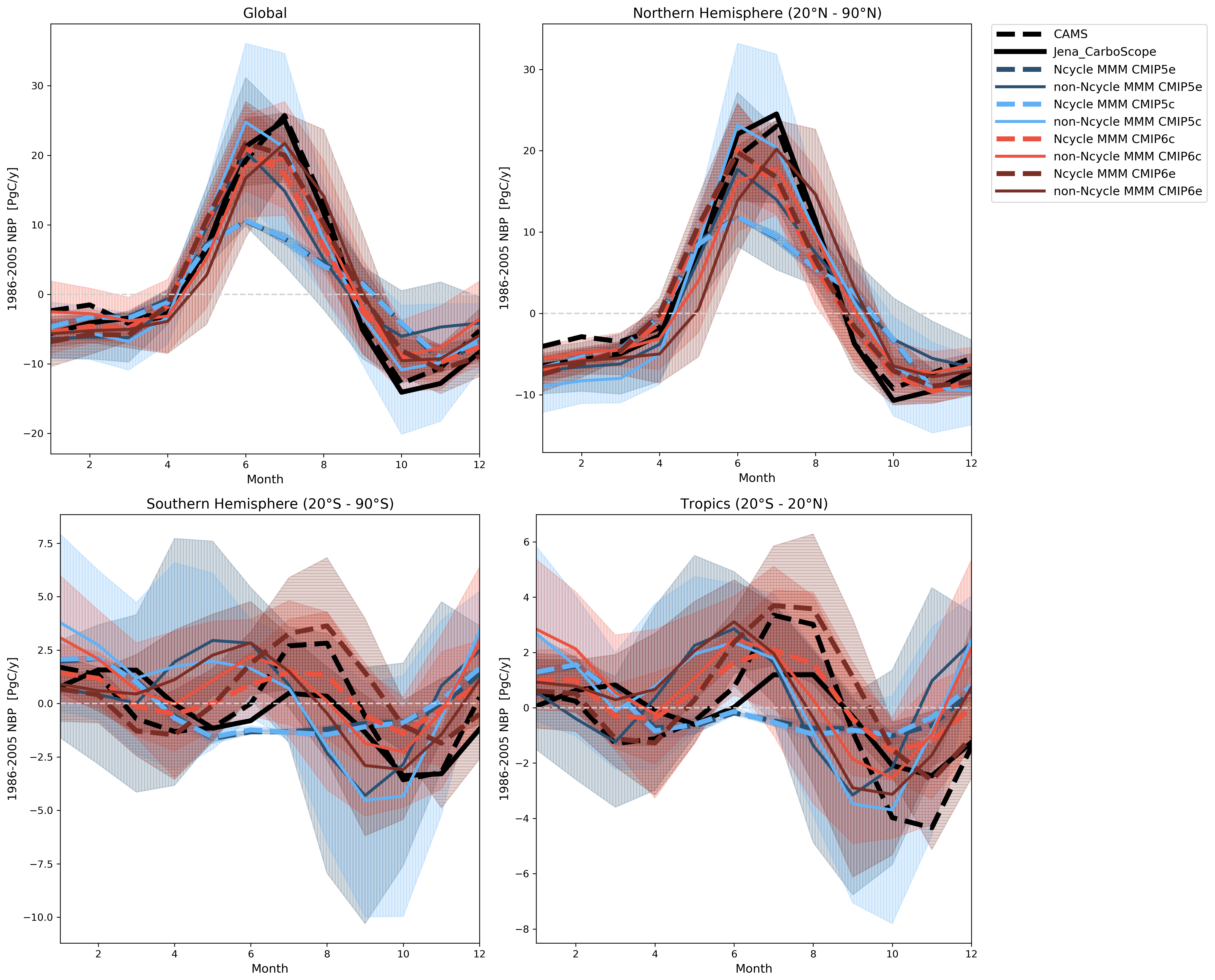} %originally width=12cm
\caption{As Figure \ref{fig:lai_cycle_allproj} but for land-atmosphere carbon flux and the reference data sets CAMS and CarboScope.}
\label{fig:nbp_cycle_proj}
\end{figure*}

The global seasonal cycle for NBP (Fig. \ref{fig:nbp_cycle_proj}) is dominated by the northern hemisphere, with almost no discernible cycle in the southern hemisphere or the tropics. There is generally a good agreement between the two inversions, but the CAMS inversions shows a larger seasonal cycle in the SH and tropics of approximately 6\,PgC yr$^{-1}$, where CarboScope shows no clear seasonal cycle. In the NH, and due to its large contribution to the total also globally, CAMS has a higher NBP at the start of the year and to a lesser degree at the end of the year, where CarboScope shows a larger negative NBP and thus carbon sink. The models agree with the carbon sink of CarboScope in these months, but have a weaker carbon sink (higher NBP) in NH autumn. The CMIP5 Ncycle MMMs have a smaller seasonal cycle amplitude compared to any of the other MMMs and the reference data, carried over from GPP. The other MMMs reproduce the seasonal cycle well, while the non-Ncycle CMIP6e MMM is shifted late by a month, showing possible issues with the start and end of the growing season. In the SH and tropics, where CarboScope found no significant cycle and CAMS had a slightly larger one in the tropics, the Ncycle models follow the shape and timing of the CAMS data, while the non-Ncycle models have a seasonal cycle shifted to two months earlier. There is no significant difference between CMIP5 and CMIP6 nor between c and e experiments in the MMM.

\begin{figure*}[t]
\includegraphics[width=18cm]{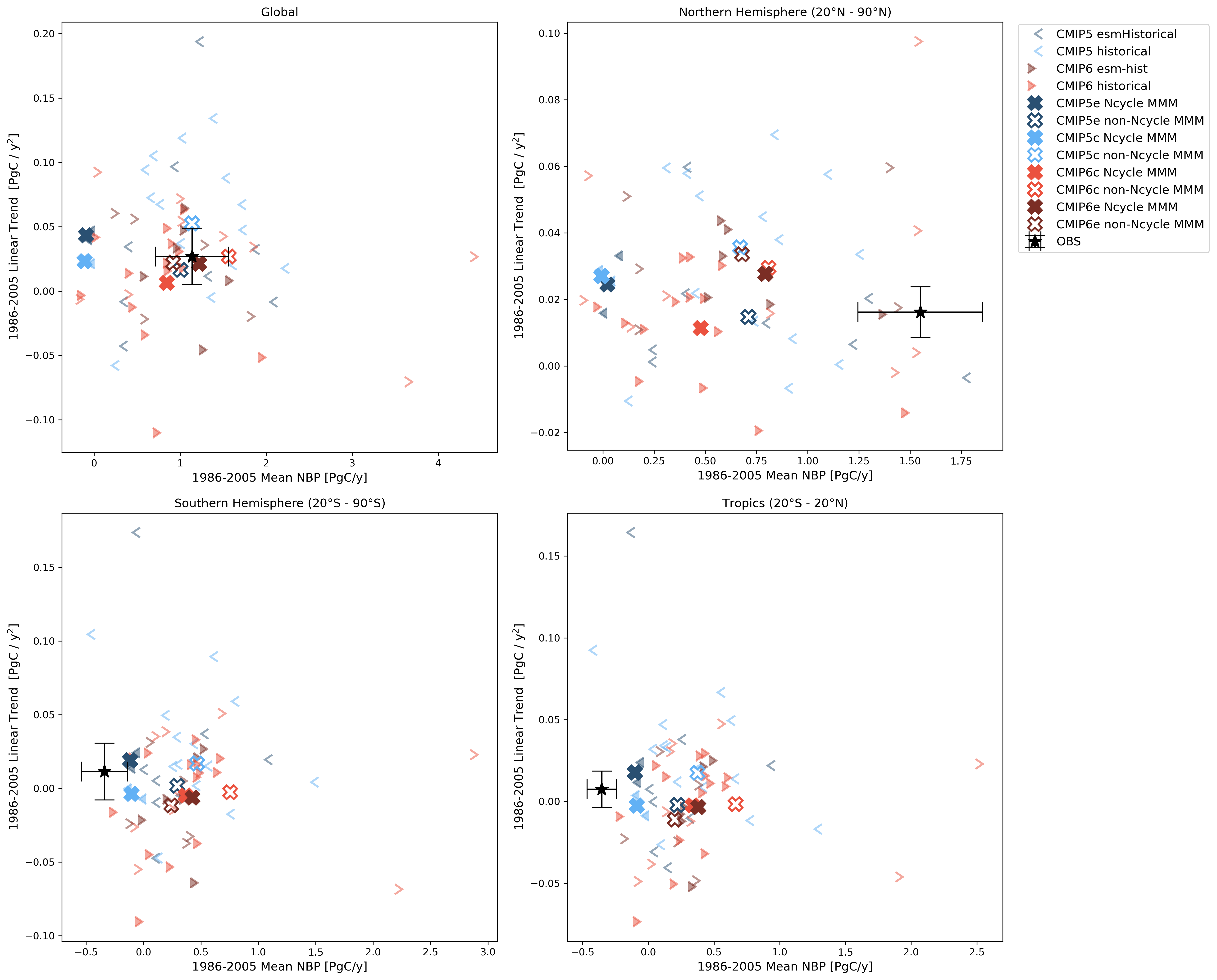} %originally width=12cm
\caption{As Figure \ref{fig:lai_scatter} but for land-atmosphere carbon flux and the reference data sets CAMS and CarboScope. The GCP data is a globally averaged timeseries and thus only appears in the global plot.}
\label{fig:nbp_scatter}
\end{figure*}

The temporal mean and trend of spatially summed NBP is shown in Figure \ref{fig:nbp_scatter} with numbers given in Tables S5 and S6. CAMS shows a larger mean NBP compared to CarboScope in all regions but the tropics, which is consistent with the NBP averages from \citet{Seiler2022} using this data set and who found its NBP to be larger than comparable data sets and model results. GCP as a global average is only available as reference data set for the global panel. Globally, the reference data sets have a mean NBP of 0.71 $\pm$ 0.94\,PgC yr$^{-1}$ for CarboScope, 1.00 $\pm$ 0.84\,PgC yr$^{-1}$ for GCP and 1.72 $\pm$ 0.94\,PgC yr$^{-1}$ for CAMS. The models show a far larger range with outliers for the INM-CM CMIP6c models. The Ncycle CMIP5 means show negative trends, while the other MMMs range between 0.84 $\pm$ 0.34\,PgC yr$^{-1}$ (Ncycle CMIP6c) and 1.56 $\pm$ 0.62\,PgC yr$^{-1}$ (non-Ncycle CMIP6c), with Ncycle MMMs showing significantly smaller mean NBPs similar to the GCP reference, while the non-Ncycle MMMs have a larger NBP but still smaller than the CAMS data. The relatively good overall agreement of the models' mean NBP with the reference data does not hold for the different regions. Most models and all MMMs simulate a lower carbon sink in the northern hemisphere when compared to the inversions, with Ncycle models generally showing a smaller mean NBP, but no large discernible differences between the different groupings. Conversely, while the inversions estimate both the southern hemisphere as well as the tropics to be a slight carbon source due to deforestation, the MMMs with the exception of the CMIP5 Ncycle show a carbon uptake by land in these regions. The large values for the non-Ncycle CMIP6c MMM are again due to the overestimation of the INM-CM4-8 and INM-CM5-0 models, but as their mean NBP in the NH is not a large ourlier, we did not remove these from the MMM. The underestimation in the NH combined with the overestimation in the SH and tropics leads to the good global agreement of the total carbon sink. This is in agreement with the findings from IPCC AR6 \citep{ipcc6ch3, ipcc6ch5}. The inclusion of a nitrogen cycle and therefore the inclusion of nitrogen limitations on CO$_{2}$ fertilization was expected to address this discrepancy of the distribution of the carbon sinks \citep{ipcc6ch5}, but the data does not support this, as the Ncycle MMMs do not show a different performance to the non-Ncycle MMMs in CMIP6.
While the models show a large range of trends, the MMMs agree well with the reference data, and this continues in the other regions as well. While N-limitation is not expected to be substantial at present day, it represents a major limitation on future land-carbon uptake \citep{Zaehle2015}, and thus its inclusion a major advance in being able to robustly simulate future carbon balance of the terrestrial carbon cycle. 

\begin{figure*}[t]
\includegraphics[width=18cm]{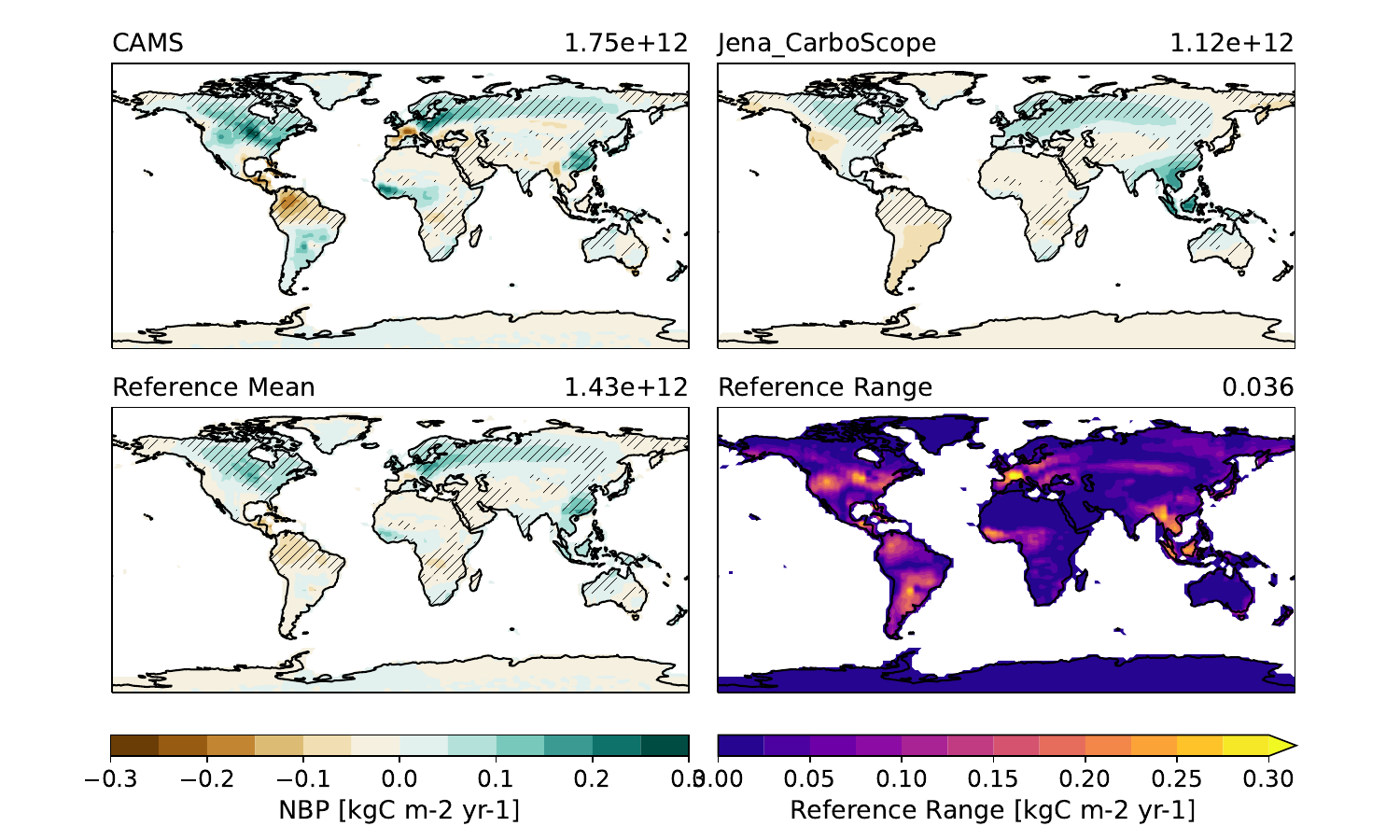} %originally width=12cm
\caption{Similar to \ref{fig:gpp_mapsref} but for land-atmosphere carbon flux. Additionally, the forward slash hatching symbolizes areas where the reference data sets agree on the sign or where the difference is smaller than the size of one bin of the contour plot.}
\label{fig:nbp_mapsref}
\end{figure*}

\begin{figure*}[t]
\includegraphics[width=18cm]{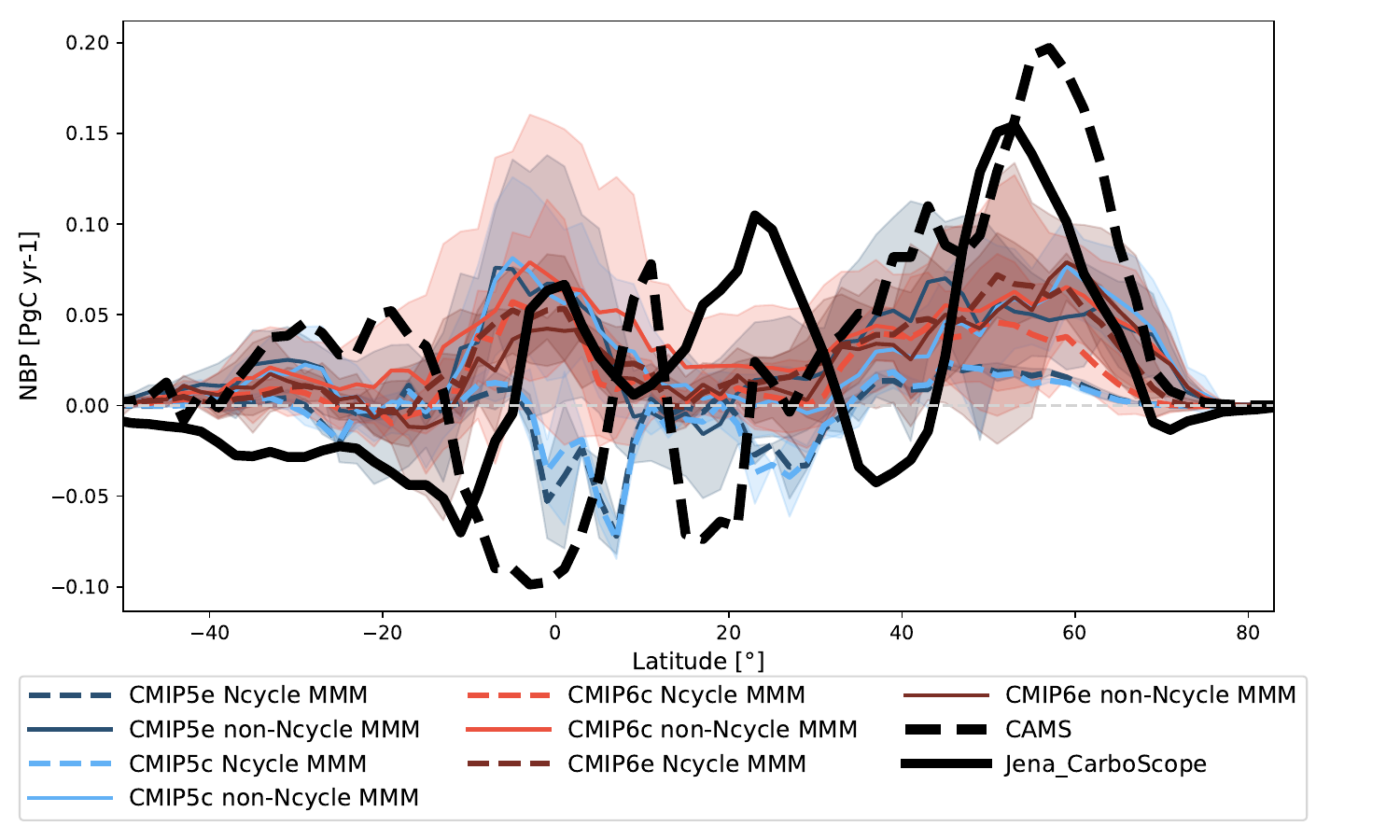} %originally width=12cm
\caption{As Figure \ref{fig:gpp_zmeans} but for land-atmosphere carbon flux with CAMS and CarboScope reference data.}
\label{fig:nbp_zmeans}
\end{figure*}

As seen before, the reference data sets show different means, trends and slightly different seasonal cycles in the different regions. For a more detailed look, Figure \ref{fig:nbp_mapsref} shows maps of the reference data. The hatched area is the area where the data sets agree on the sign or within a margin of half the bin size of the contour plot. While in large parts of the globe the data sets agree in sign, there are significant differences. In North America CAMS shows a much larger carbon sink than CarboScope which instead shows some carbon sources along the west coast and throughout South America. In CAMS South America is split into a much stronger carbon source in the amazonian rainforests and a carbon sink in the southern part. The data sets also disagree in Europe, which is a carbon source according to CAMS but a sink in CarboScope. Literature found Europe to be a carbon sink for the first part of the 21st century \citep{IPCC_2013_WGI_Ch_6, Reuter2014} using different reference data sets. This would support the CarboScope data set but due to the different time frames considered it is not definitive. CAMS also sees a carbon sink in tropical Africa where CarboScope has a slight carbon source. In South East Asia, CarboScope has a large carbon sink, where CAMS shows a neutral carbon flux. The area of most agreement is in the NH as a large carbon sink where there is no deforestation and a bit of aforestation. The amazon region was a large carbon sink which is becoming a source due to deforestation \citep{Gatti2021}. CAMS sees the amazon as a strong carbon source, while CarboScope shows a smaller source, with a sink in the northwestern region. \citet{Keenan2018} found inverse models to show south America as both a carbon source, as well as a carbon sink and is thus a hotly debated area. \citet{Kou-Giesbrecht2023N} attribute the weak agreement between CarboScope and CAMS to differences in the inversion models and atmospheric CO$_2$ measurements used, with larger differences at latitudes with smaller land areas. Due to the difference between the observational data sets, the bias maps to a reference mean shown for the other considered variables so far have been omitted. Instead, the area weighted zonal sums are plotted in Figure \ref{fig:nbp_zmeans} for comparison of the MMMs with both reference data sets. The reference data sets disagree for almost all latitudes, thus making the model comparison to the reference data in these regions not very meaningful. The area where the reference data is in most agreement is in the northern high latitudes (50--80\,$^\circ$N), where both of them show a strong carbon sink, about double of that shown in both CMIP5 and CMIP6 models. While the issues in CLM4 and thus the CMIP5 Ncycle models are clearly visible (similarly to LAI and GPP), the other MMMs show similar NBP in all latitudes. 

\subsection{Carbon Stocks}

\begin{figure*}[t]
\includegraphics[width=14cm]{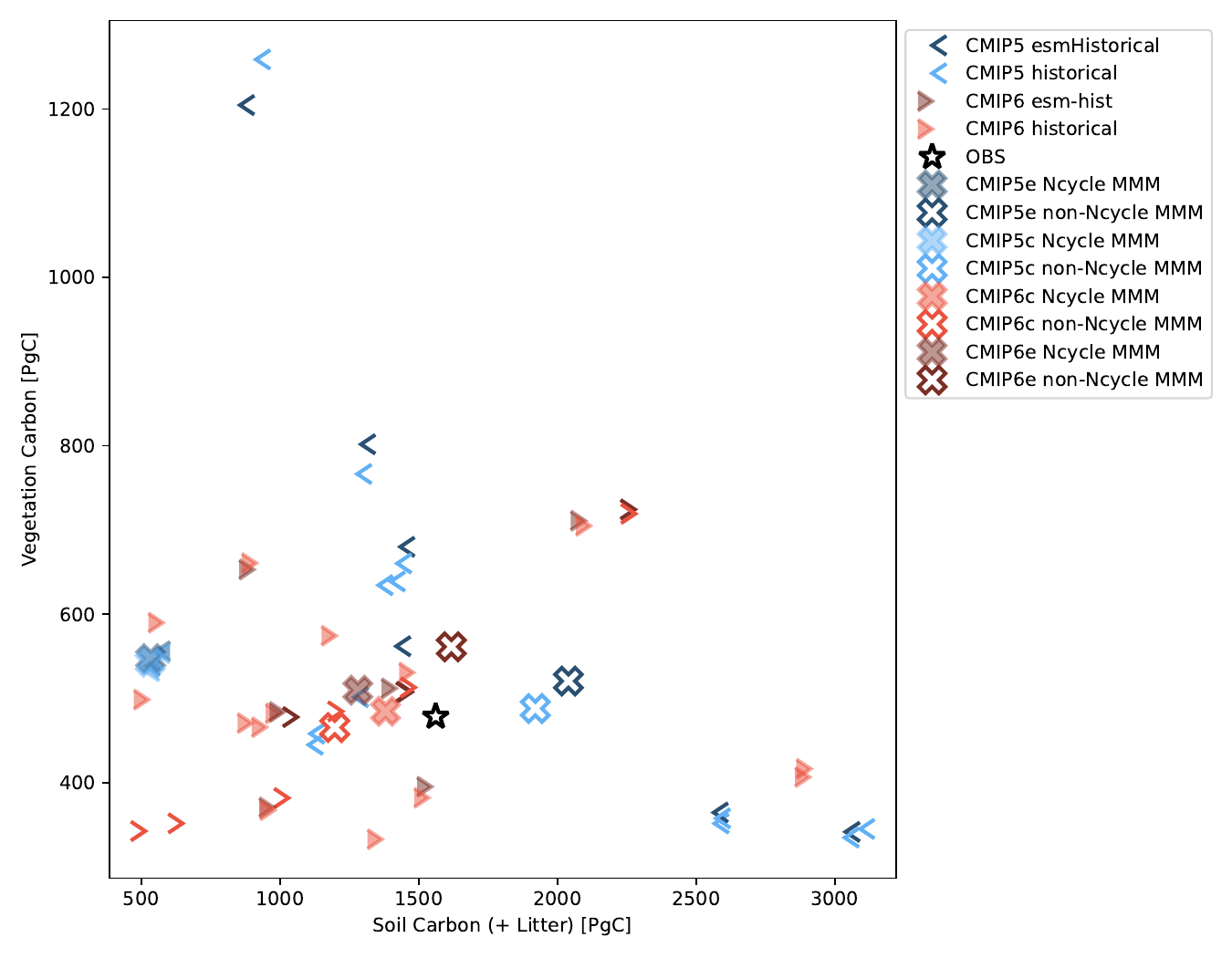} %originally width=12cm
\caption{Scatter plot of global mean vegetation and soil carbon over 1986-2005, with observations from NDP (vegetation carbon) and HWSD+NCSCD (soil carbon). As in Figure \ref{fig:lai_scatter}, filled symbols denote models with nitrogen cycle.}
\label{fig:csoil_cveg}
\end{figure*}

Another large uncertainty in CMIP5 was the amount of carbon stored in soil and vegetation. This leads to large uncertainties in land-use change emissions which are important for quantifying cumulative emissions as well as climate mitigation strategies \citep{Friedlingstein2023}. \citet{Varney2023} investigated the carbon-climate feedbacks of soil and vegetation carbon and found soil carbon to be the dominant response of the land surface, highlighting the need to reduce the uncertainty in carbon storage to better quantify future changes of the climate system. Figure \ref{fig:csoil_cveg} shows a scatterplot of the global sums vegetation against the combined soil and litter carbon. The observational soil (HWSD+NCSCD) and vegetation carbon (NDP) data sets are derived from in situ measurements taken over a long period of time and are thus given without a time coordinate, while the models were averaged over 1986-2005. Note that some models (CanESM5-CanOE CMIP6c, GFDL-ESM4 CMIP6e, INM-CM4-8 CMIP6c, INM-CM5-0 CMIP6c, FIO-ESM CMIP5e, CanESM2 CMIP5c and inmcm4 CMIP5c) did not have data on the ESGF nodes for soil or vegetation carbon and are thus missing from the carbon stocks analysis.  BNU-ESM CMIP5c and CMIP5e shows a far larger vegetation carbon than the other models and is thus removed in the calculation of the mean. Additionally, CLM5 and thus CESM2, CESM2-WACCM, NorESM2-LM, and NorESM2-MM include a full vertical soil profile. For these models, the \textit{cSoilAbove1m} variable is used for better comparison with the other models, as done in \citet{Varney2022}.
The large spread in the global carbon stocks still remains in CMIP6 as shown in Figure \ref{fig:csoil_cveg} with values for each data set listed in Table S7. In CMIP5 vegetation carbon was spread between 335\,PgC (MPI-ESM-LR CMIP5c) and 802\,PgC (GFDL-ESM2M CMIP5e) with the outlier of BNU-ESM even reaching values above 1200\,PgC. The spread has only marginally been reduced in CMIP6 to a range of 333\,PgC (EC-Earth3-Veg CMIP6c) to 724\,PgC (CNRM-ESM2-1 CMIP6e). The reference data are at a value of 478\,PgC, in the lower range of the models, with the MMMs ranging between 465\,PgC (non-Ncycle CMIP6c) and 547 (Ncycle CMIP5e). The spread in soil carbon is even larger with a CMIP5 range of 513\,PgC (CESM1-BGC CMIP5c) up to 3092\,PgC (MPI-ESM-MR CMIP5c). The overestimation by MPI-ESM is due to its decomposition parameterization depending on soil moisture and showing maxima in continental dry lands. In CMIP6 MPI-ESM1-2 the soil carbon model was changed to YASSO which simulates more plausible soil carbon content \citep{Mauritsen2019}. The spread in soil carbon was not significantly reduced in CMIP6 with a range of 514\,PgC (GFDL-ESM4 CMIP6c) to 2913\,PgC (CMCC-ESM2 CMIP6c), with the reference value for HWSD+NCSCD at 1561\,PgC. The CMIP5 Ncycle models have a soil carbon on the lower end of the range, consistent with the CMIP6 models TaiESM1 and SAM0-UNICON which also use CLM4. The CMIP5 Ncycle MMMs are on the very low end of the range with 532\,PgC and 534\,PgC for CMIP5 and CMIP5e respectively, while the other MMMs are closer to the reference data and range between 1197\,PgC for non-Ncycle CMIP6c and 2040\,PgC for non-Ncycle CMIP5e. While the CMIP6 Ncycle MMMs are closer to the reference data, no significant improvement due to the inclusion of the interactive nitrogen cycle can be seen when considering the whole spread of the models. This is consistent with \citet{Wang2022} who found that changing models from C to CN coupling often result in lowered ecosystem storage, but due to different parametrizations simulate similar carbon pools. \citet{Varney2022} suggest that much of the uncertainty in carbon stocks is due to the simulation of below-ground processes - this is backed up by the differences in soil carbon being much greater than in GPP, and thus implicating differences in simulated residence times \citep{Carvalhais2014, Todd-Brown2014}. For a more in-depth discussion on we would like to refer to dedicated studies, such as \citet{Varney2022} and \citet{Wei2022}. Furthermore, \citet{Varney2023priming} found that while the CMIP6 future soil carbon projections have a lower model spread compared to CMIP5, the structure of soil carbon models within CMIP6 ESMs has likely contributed towards this reduction.

\subsection{Overall Model Performance}

\begin{figure*}[t]
\includegraphics[width=18cm]{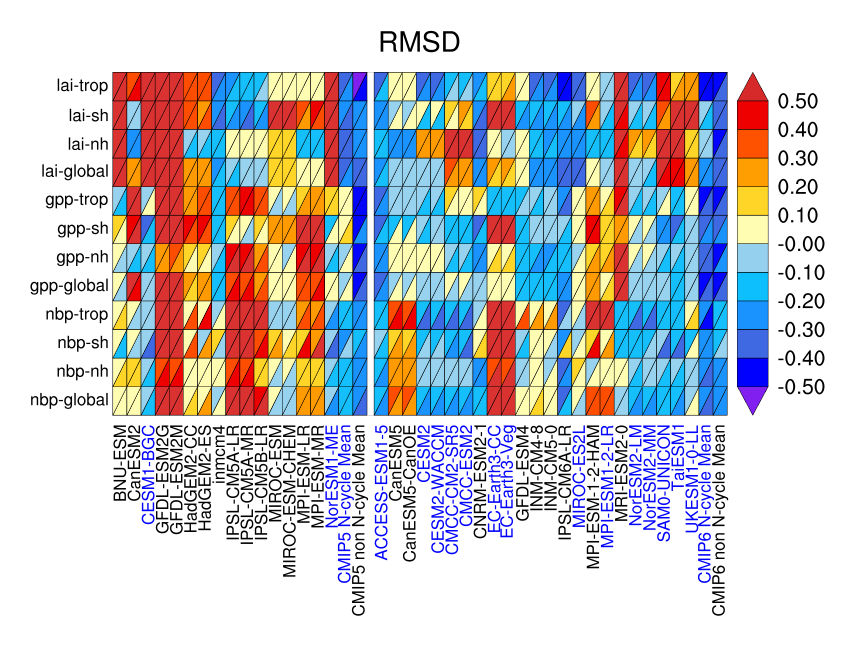} %originally width=12cm
\caption{Relative space-time root-mean-square deviation (RMSD) performance metrics for CMIP5 (left) and CMIP6 (right) concentration driven simulations for variables relevant to the carbon cycle compared to reference data sets. Blue shading indicates a performance better than the median RMSD of all models in the plot, while the redder the color, the worse the performance. The RMSD is normalized relative to the median of all models. The considered time periods depend on the start of the observational data (see Table \ref{tab:obs}) and end in 2005 to accommodate the end of the CMIP5 data. When using two observational references, a diagonal split is introduced, with the default reference data set being shown on the lower right, while the alternate data set is used for the top left triangle. The default and alternate reference data sets are marked in Table \ref{tab:obs} and are as follows: LAI: LAI4g (main), GLASS (alt); GPP: FLUXCOM (main), GLASS(alt); NBP: CarboScope (main), CAMS (alt).}
\label{fig:hist_perf}
\end{figure*}

\begin{figure*}[t]
\includegraphics[width=14cm]{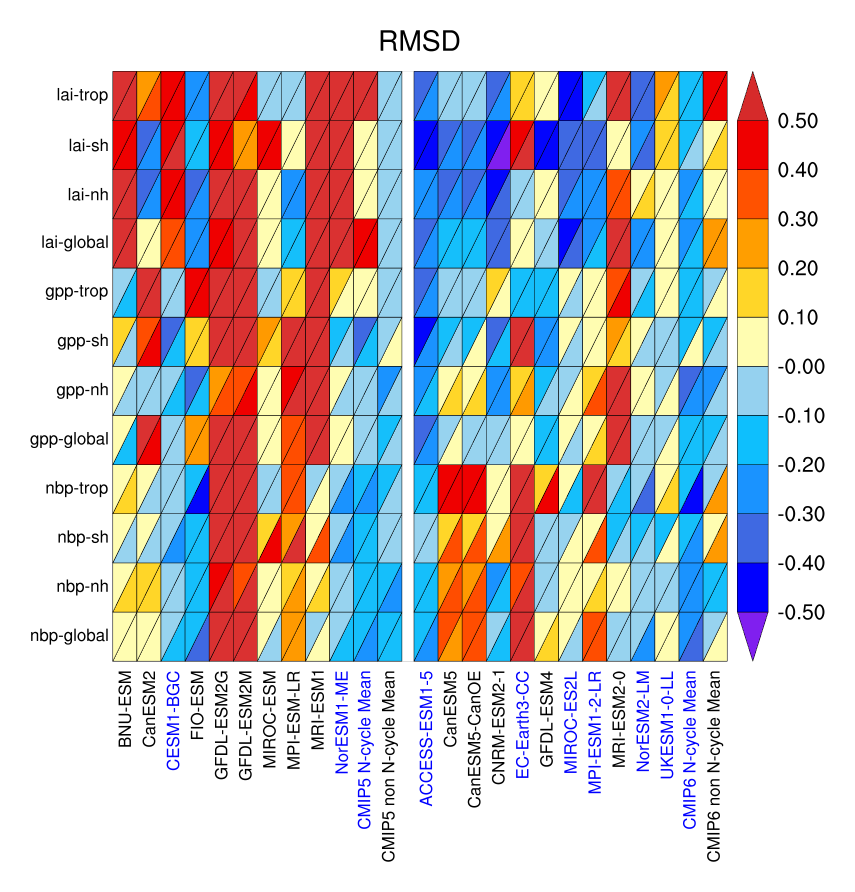} %originally width=12cm
\caption{As Figure \ref{fig:hist_perf} but for emission driven simulations.}
\label{fig:esmhist_perf}
\end{figure*}

In this Section we assess the overall performance of CMIP5 and CMIP6 models with respect to carbon cycle variables. Figure 18 shows a performance metrics (portrait) plot similar to \citet{Gleckler2008}. It is produced by calculating the normalized relative space-time root mean square difference (RMSD) of the climatological seasonal cycle of a model variable with respect to a reference observation. The normalization is done relative to the ensemble median of both CMIP5 and CMIP6 models, with positive values (red) denoting a higher RMSD and thus worse performance while negative values (blue) denote a lower RMSD than the ensemble median and thus a better performance. As the carbon stocks from the observations do not vary in time, the calculation of the RMSD as done here is not meaningful and thus only NBP, GPP and LAI are shown in the plot for all four considered regions (global, northern hemisphere, southern hemisphere, tropics). MMMs for both Ncycle and non-Ncycle models were added, with the models which were excluded from MMMs due to various issues as stated in the previous sections also removed from the MMM here. Note that the MMMs were calculated on the climatologies prior to calculation of the RMSDs, so over- and underestimations can cancel each other out. This is the standard for the performance metrics plot implemented in ESMValTool and kept for consistency. Variables with two reference data sets show the main reference in the lower right triangle, while the alternate reference is shown for the upper left triangle. data sets marked in bold in Table \ref{tab:obs} are the main references. CMIP5 models are shown on the left and CMIP6 models on the right, with figures for both concentration driven (Fig. \ref{fig:hist_perf}) and emission driven models (Fig. \ref{fig:esmhist_perf}). Models with a nitrogen cycle are marked with blue labels.

Most models have similar scores when compared to the different observations for GPP and LAI, showing that the inter-model spread is CMIP6 is larger than the observational uncertainty in these variables. For NBP however, models can have different scores to the considered reference data, which is due to the difference in the reference data found in the previous sections. Models on average perform much better than CMIP5 models, with models that had a predecessor in CMIP5 improving on their CMIP5 performance in almost all variables, such as GFDL and IPSL in all variables and CESM and NorESM in LAI, with the exception of CanESM which shows a reduced performance for NBP in CMIP6. Large improvements can be found in all variables going from CMIP5 to CMIP6, especially in LAI with the exception of the models using the older CLM4 land component (SAM0-UNICON and TaiESM1) in CMIP6 and GPP, which were previously identified as weaknesses in CMIP5. Only the MRI-ESM2-0 model shows a bad performance in both these variables. As mentioned before, dynamic vegetation in models plays a large role in their ability to simulate variables directly related to it, with models interactively simulating vegetation cover (marked with a D in Tables \ref{tab:CMIP6} and \ref{tab:CMIP5}), showing a below average score for LAI and GPP RMSD.

Models which were remarked upon in the previous sections as having good or bad agreement with the observations in specific areas, such as the NBP problems in December for EC-Earth3 have RMSDs that reflect these statements, making this metric a well-suited measure for overall performance. Most models have similar RMSDs in the different regions, with the global value reflecting a mean of the different regions. There does not seem to be a qualitative difference between Ncycle and non-Ncycle models as a whole, but the MMMs perform better than any individual model. The better performance of the MMMs is mathematically expected as long as the assumption that both observations and model simulations draw from the same distribution holds true \citep{Christiansen2018}. The global NBP, LAI, and GPP are also found in Figure 42 of chapter 3 of the IPCC AR6 \citep{ipcc6ch3}, which showed not only carbon cycle variables but also other land, ocean and atmosphere variables averaged over 1980-1999 for comparison across models from CMIP3 to CMIP6. Their results are compared to reference data from JMA-TRANSCOM for NBP, LAI3g for LAI, and MTE and FLUXCOM for GPP. As other than NBP these reference sets are the same as the ones considered in this paper, the results are also the same. For NBP despite the different data set, the performance of the models is very similar to the one found for CAMS, our alternative data set. The ILAMB benchmark used in chapter 5 of the IPCC AR6 \citep{ipcc6ch5} also comes to the conclusion of model improvement from CMIP5 to CMIP6.
No qualitative difference can be found between models that have both emission driven and concentration driven simulations compared to models with only concentration driven simulations, and models with both simulations have similar RMSDs in both. This indicates that carbon exchanges are well simulated in these models as the freely evolving fluxes are comparable to results with prescribed atmospheric concentrations.

\begin{figure*}[t]
\includegraphics[width=18cm]{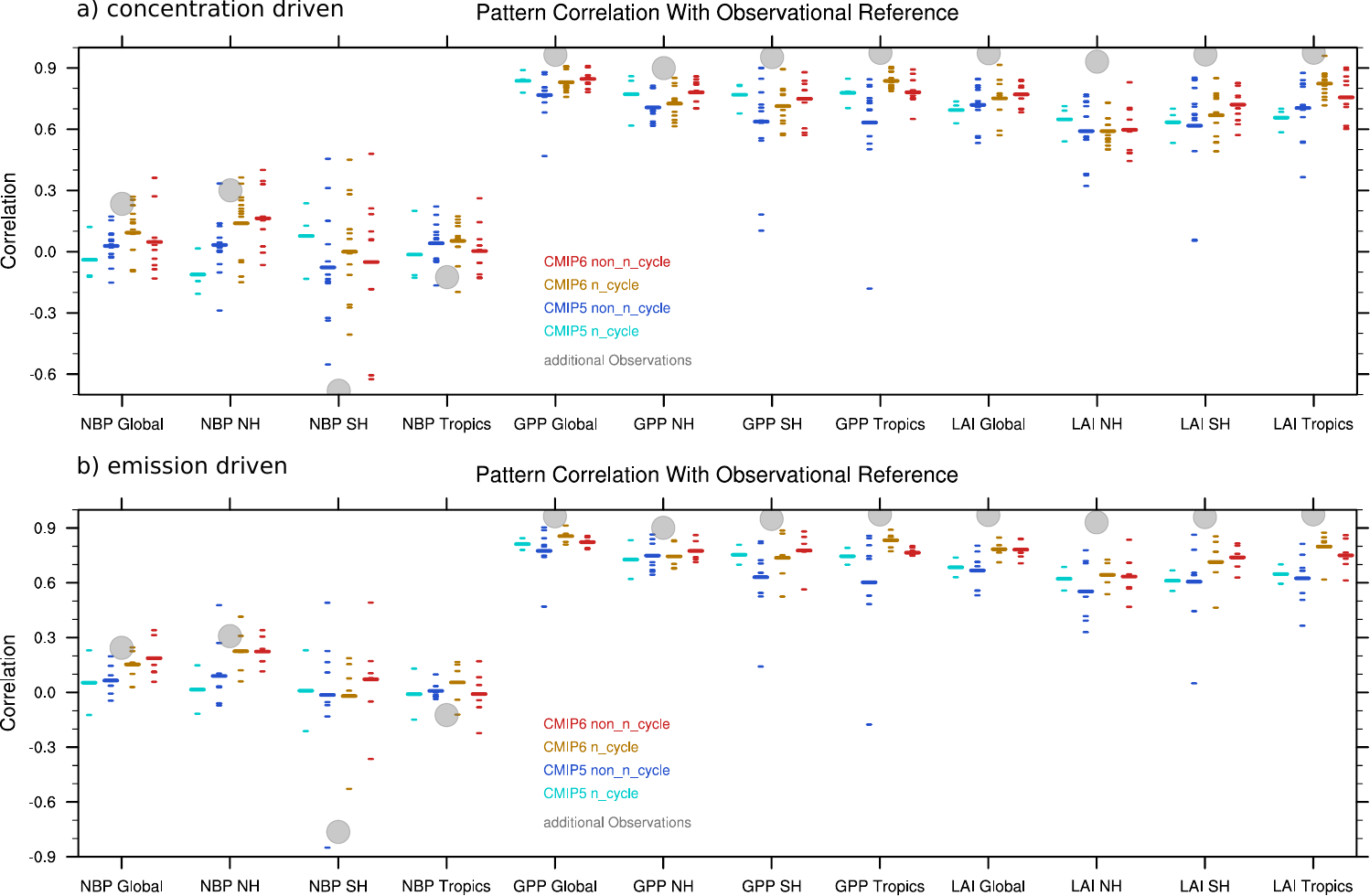} %originally width=12cm
\caption{Centered pattern correlations between models and reference data sets for annual mean climatology for concentration driven (a) and emission driven (b) CMIP5 and CMIP6 models, split into Ncycle and non-Ncycle models. Main and alternate observations are the same as in Figure \ref{fig:hist_perf}.}
\label{fig:patterncor}
\end{figure*}

Centered pattern correlations for these variables and regions are shown in Figure \ref{fig:patterncor}, with a score of 1 meaning perfect similarity of a model to the reference data, while a value of 0 signifies no relationship. The longer lines denote the MMM, while the grey circle shows the similarity of the alternate data set to the main reference data set. For GPP and LAI the reference data sets show very good similarity of above 0.9, while for NBP the differences of the references highlighted in Figure \ref{fig:nbp_mapsref} is highlighted through a small correlation of up to 0.3, with a high anti-correlation in the southern hemisphere. Due to this, the precise value of the correlation coefficient between models and reference data set is not a good measure, but it can be seen that the models show a large spread. For GPP, the CMIP6 performance in the tropics is vastly improved, with even higher correlation values for Ncycle models. Other than in the NH, the CMIP5 models show a large spread in correlation values, which has reduced for CMIP6. The correlation distribution for LAI is similar as GPP, with the highest correlation values found in the tropics and globally, but the difference between Ncycle and non-Ncycle models is not as prominent. These overall performance plots underline the specific conclusions from the separate sections above.

\conclusions[Summary and Conclusion]  %% \conclusions[modified heading if necessary]

To be able to have confidence in model projections of climate change, Earth System Models first need to show the ability to simulate observed climatologies and trends of the carbon cycle in the present day climate. In the Coupled Model Intercomparison Project Phase 5 \citep[CMIP5, ][]{Taylor2012}, several weaknesses of the simulated carbon cycle were found, such as a general overestimation of photosynthesis and a wide range of values for carbon stocks, which became one of the main areas of focus for improvement for some model groups \citep{Delire2020}. In this study, we have analysed the land carbon cycle of models participating in the Coupled Model Intercomparison Project Phase 6 \citep[CMIP6,][]{EyringCMIP6} to investigate whether these weaknesses were improved in the newer model generation, with a special focus on differences arising due to inclusion of an interactive terrestrial nitrogen cycle in some of the CMIP6 models. Concentration and emission driven simulations from CMIP5 and CMIP6 models were compared to reference data sets, with 2 out of 18 CMIP5 models and 15 out of 23 CMIP6 models including carbon-nitrogen interactions. We assessed means, trends and seasonal cycles of the leaf area index (LAI), the gross primary productivity (GPP), and the land-atmosphere carbon flux (NBP). We furthermore looked at land carbon stocks to see if the large range of values simulated in CMIP5 was reduced in CMIP6.

In general, CMIP6 models show a better performance across all assessed land carbon cycle variables to differing degrees, and no significant differences between the concentration driven and emission driven simulation were found in the considered variables, that cannot be explained by the number of different models. While there is a bias towards the CLM land component in the CMIP6 models, the different versions (4, 4.5, 5) do not perform the same and thus these versions can be seen as independent components for the multi-model mean.

The leaf area index was a weakness of the CMIP5 simulation, as its seasonal cycle was not well captured and its absolute value was generally overestimated. While the peak of the climatological seasonal cycle of LAI is much better reproduced in CMIP6, the amplitude of the seasonal cycle is weaker in CMIP6 compared to observations due to a weaker drawdown in winter. Thus LAI should remain an area of focus for future model development. Mean LAI is much better reproduced in CMIP6, while the range of trends in the observations is large enough to cover most models for both CMIP5 and CMIP6. It should be noted that due to correlations between parameters, there are often tradeoffs for better reproducing one variable. In CLM5 such a tradeoff had to be weighed between biases for GPP and LAI against high plant functional type (PFT) survivability rates (Lawrence et al., 2019). Therefore, looking at one variable separately instead of the whole model performance can lead to wrong conclusions about the model's ability of reproducing the carbon cycle, depending on which choices were made in the tuning. Similarly, models interactively simulating vegetation cover perform worse in the evaluation of present-day LAI compared to models using observationally derived landcover maps due to simulating trees and grasses in the wrong areas. However, only these models with dynamic vegetation can account for future changes in vegetation and the impact of these changes on climate and carbon processes in future projections.

One of the largest improvements due to the inclusion of an interactive nitrogen cycle was seen in GPP, where the CMIP6 nitrogen cycle models were able to capture the seasonal cycle in the northern hemisphere well, which was previous overestimated. Beside the improvements in the NH, bias patterns in the tropics showing larger GPP overestimations bordering tropical rainforests are reduced in CMIP6 models, with some of these biases wholly removed in the multi-model mean of the CMIP6 models with interactive nitrogen cycle.

The land carbon sink is underestimated in the northern hemisphere regardless of CMIP phase or inclusion of nitrogen cycle. The models compensate for this by simulating a larger carbon sink in the tropics and the southern hemisphere for a global average close to the observed value. An improvement is seen in CMIP6 in capturing the amplitude of the seasonal cycle, which is controlled by carbon uptake through photosynthesis in the growth season and carbon release by respiration. This improvement can largely be attributed to the improved seasonal cycle of GPP.

The large range of soil and vegetation carbon was another large weakness of CMIP5, with inter-model differences of 900 PgC for vegetation carbon and 2500 PgC for soil carbon. This range has not significantly decreased in CMIP6, and it remains an area for improvement.

While we find a significant improvement of the inclusion of the nitrogen cycle for photosynthesis, the effects are reduced for the leaf area index and the land-atmosphere carbon flux. Despite similar NBP for models with and without interactive nitrogen cycle, models without interactive nitrogen overestimate carbon fertilization, leading to large differences of atmospheric carbon content for future scenario simulations \citep{Kou-Giesbrecht2023}. Therefore, the inclusion of further limiting nutrients like phosphorus is important, as they will likely have substantial impacts on future carbon uptake \citep{Yang2023}. Model performance overall has improved from CMIP5 to CMIP6 even with the added complexity introducing more degrees of freedom into the models, as also found in the latest IPCC report \citep{ipcc6ch3, ipcc6ch5}. This is a positive outlook for the future, as many aspects have to be considered when increasing model complexity, such as a need to adjust existing parametrisations after model structural changes from carbon-only to carbon-nitrogen coupling. Without such adjustments, lowered ecosystem carbon storage simulated by models with N processes would lead to an underestimation of carbon pools \citep{Wang2022}. The increased overall model performance confirms results from the individual model groups who found improved performance in carbon cycle variables compared to previous model configurations, with the biggest improvements seen in LAI and GPP \citep{Ziehn2017, Danabasoglu2020}. Many areas requiring improvement remain, such as simulated carbon stocks which saw no significant reduction in the simulated range between CMIP5 and CMIP6, or the inclusion of more nutrient limitations like an interactive phosphorus cycle. The improvement of the carbon cycle in the models since CMIP5 is a step in the right direction for a better understanding and a more accurate simulation of future trends. Based on our analysis, due to the small differences between historical concentration and emission driven simulations despite the increased process-realism, we recommend ESMs in future CMIP phases to be based on emission driven simulations to fully account for climate-carbon feedbacks in future projections, supporting the message from \citet{Sanderson2023}. Similarly, due the significant improvements in GPP with the inclusion of an interactive nitrogen cycle and no detrimental change in the present day evaluation of any carbon cycle variable, we suggest that the nitrogen cycle should be seen as a necessary part of carbon cycle models in the future.

%% The following commands are for the statements about the availability of data sets and/or software code corresponding to the manuscript.
%% It is strongly recommended to make use of these sections in case data sets and/or software code have been part of your research the article is based on.

%\codeavailability{} %% use this section when having only software code available

%\dataavailability{} %% use this section when having only data sets available

\codedataavailability{The code to reproduce this study is part of ESMValTool \citep{Righi2020, Eyring2020}. The corresponding recipes can be found under the folder gier23bg. ESMValTool v2 is released under the Apache License, version 2.0. The latest release of ESMValTool v2 is publicly available on Zenodo at \url{https://doi.org/10.5281/zenodo.3401363} \citep{Zenodo_ESMValTool}. The source code of the ESMValCore package, which is installed as a dependency of ESMValTool v2, is also publicly available on Zenodo at \url{https://doi.org/10.5281/zenodo.3387139} \citep{Zenodo_ESMValCore}. ESMValTool and ESMValCore are developed on the GitHub repositories available at \url{https://github.com/ESMValGroup} (last access: 26 January 2024). CMIP data is available for download via ESGF nodes (https://esgf-node.llnl.gov/search/cmip5/, https://esgf-node.llnl.gov/search/cmip6/). Reference data sets have been formatted for use with ESMValTool using so-called CMORizers. DOIs for all data sets contributing to this study are listed in Tables 1-3 along with their references as available.} %% use this section when having data sets and software code available

%\sampleavailability{TEXT} %% use this section when having geoscientific samples available

%\videosupplement{TEXT} %% use this section when having video supplements available

%\appendix
%\section{}    %% Appendix A

%\subsection{}     %% Appendix A1, A2, etc.

\appendix

\section{Carbon Cycle in the CMIP models}
While a summary of the model components for each model can be found in Tables \ref{tab:CMIP6} and \ref{tab:CMIP5}, some special characteristics of the carbon cycle in the land surface components are given below. As several ESMs use the same land model, the sections are listed by land model instead of ESM. If different versions of the same component were used in different CMIP models, the difference between the versions, as well as the models using each version are explained.

\subsection{CABLE + CASA-CNP}
The Commmunity Atmosphere-Biosphere Land Exchange model \citep[CABLE,][]{Kowalczyk2013} version 2.4 is a land surface model coupled to the biogeochemistry module Carnegie-Ames-Stanfoard Approach carbon cycle model with nitrogen and phosphorus cycles \citep[CASA-CNP,][]{Wang2010} used in ACCESS-ESM1-5 \citep{Ziehn2020}. CASA-CNP and thus ACCESS-ESM1.5 is the only model in this study to include a phosphorus cycle coupled to the land carbon-nitrogen cycle. A sensitivity study of allowable emissions to nutrient limitation found a reduction of the land carbon uptake by 35-40\,\% with nitrogen limitation and a further 20-30\,\% reduction with nitrogen and phosphorus limitation on the carbon cycle \citep{Zhang2013}, showing the importance of nutrient limitation.

A simple land-use scheme accounts for annual net change in the vegetation tile fractions of each grid-cell which consider 10 vegetated and three non-vegetated surfaces. Three live and six dead carbon pools are modelled. Leaf area index (LAI) in ACCESS-ESM1-5 is calculated from specific leaf area and the size of the leaf carbon pool, while phenology is prescribed. In the CMIP5 model ACCESS-ESM1, which is not considered in this paper due to a lack of variables on the ESGF, LAI was significantly higher than observations, mainly due to an overestimation of LAI in the northern hemisphere, despite a significant underestimation of LAI in the tropics. To better match the observations, two parameters were adjusted for ACCESS-ESM1-5: one PFT-specific parameter used in the parametrisation for the maximum carboxylation rate and thus related to the nitrogen cycle, as well as one global parameter related to the daytime leaf respiration rate. Further changes to the model since CMIP5 include the conservation of land carbon, which was not conserved in CMIP5, as well as the inclusion of wetland tiles in the biogeochemistry calculation and the removal of a spin-up condition which ensured a minimum nitrogen and phosphorus level in soil pools.

\subsection{CLASS + CTEM}
The land component in the Canadian Earth System Models (CanESM) is divided into the physical part represented by the Canadian Land Surface Scheme \citep[CLASS, ][]{Verseghy1991, Verseghy2000, Verseghy1993} and the biogeochemical processes as simulated by the Canadian Terrestrial Ecosystem Model \citep[CTEM,][]{Arora2003,AroraBoer2003,AroraBoer2005}. In the CMIP5 model \citep[CanESM2,][]{Arora2011} version 2.7 of CLASS was used, while the CMIP6 models CanESM and CanESM-CanOE \citep{Swart2019} employ CLASS v3.6. While neither version includes a nitrogen cycle, a parameter representing terrestrial photosynthesis downregulation is included to simulate the effect of nutrient constraints. This parameter is increased in CanESM5 compared to the previous version CanESM2, resulting in a higher land carbon uptake in CanESM5. Four PFTs are considered in CLASS, while CTEM increases the number to nine PFTs so that phenology can be simulated prognostically. LAI is dynamically simulated and three live and two dead carbon pools are considered. Added features since CanESM2 include dynamic wetlands and their diagnostic methane emissions.

\subsection{CLM}
The Community Land Model \citep[CLM,][]{CLMtechnote} is the most commonly used land model in this study, with 11 models across CMIP5 and CMIP6 using 3 different versions of it. For the CMIP5 models, CLM3.5 \citep{CLM3.5_Oleson2008} was used by FIO-ESM and CLM4 \citep{CLM4_Lawrence2011} was used in CESM1-BGC and NorESM1-ME. In CMIP6 CLM4 is used for SAM0-UNICON and TaiESM who mainly adapted the CESM1 configuration \citep{Lee2020}. CMCC-CM2-SR5 and CMCC-ESM2 use CLM4.5 \citep{CLM4.5_Koven2013}, while the newest version CLM5 \citep{CLM5_Lawrence2019} is a part of CESM2, CESM2-WACCM, NorESM2-LM and NorESM2-MM.

While in CLM3.5 nitrogen limitation was merely represented by a downregulation factor, CLM4 introduced the coupled carbon-nitrogen cycle. Further improvements in CLM4 included transient land cover change modeling, changes to the PFT distribution and more realistic modeling of permafrost regions. To reduce biases found in CLM4 such as low soil carbon stocks and unrealistic values for GPP and LAI in several regions such as a stark overestimation in the tropics, several parametrisations were changed in CLM4.5. Modifications were made to the canopy processes, including co-limitations on photosynthesis and photosynthetic parameters. Newly introduced features included a vertically resolved soil biogeochemistry with vertical mixing of soil carbon and nitrogen and a more realistic distribution of biological fixation over the year. The structure of the litter and soil carbon and nitrogen pools was adapted to the Century model and $^{13}$C and $^{14}$C carbon isotopes were introduced.

Finally in CLM5 many major components of the land model were updated, with a focus on a better representation of land use and land-cover change as well as a more mechanistic treatment of key processes. Changes included a stronger soil moisture control on decomposition, the use of $^{13}$C and $^{14}$C isotopes for crops, and several changes to the nitrogen cycle and its impact on photosynthesis. Flexible plant C:N ratios were introduced to eliminate instantaneous down-regulation of photosynthesis, leaf nitrogen was optimized in the form of the leaf use of nitrogen for assimilation \citep[LUNA, ][]{LUNA_Ali2016} model and a model handling the fixation and uptake of nitrogen \citep[FUN, ][]{FUN_Shi2016} was included. With respect to the land use and land cover aspect of the model, land unit weights are no longer fixed during the simulation and the transient PFT distribution was updated. CLM5 considers 22 live and 7 dead carbon pools as well as 22 PFTs.

\subsection{CoLM+BNU-DGVM}
The Common Land Model \citep[CoLM,][]{Dai2003} which shares an initial version with CLM but was then developed separately, is the land model component for the CMIP5 model BNU-ESM in the CoLM2005 version. CoLM includes a photosynthesis-stomatal conductance model for sunlit and shaded leaves separately. While carbon-nitrogen cycle interactions were included in the model, they were turned off for the CMIP5 simulations due to not being fully evaluated at the time \citep{Ji2014}.

\subsection{HAL}
The land model for the MRI models in both its CMIP5 version MRI-ESM1 and the CMIP6 version MRI-ESM2-0 is the Hydrology, Atmosphere, and Landsurface model \citep[HAL,][]{HAL_Hosaka2011}. It consists of three submodels called SiByl (vegetation) with grass and canopy vegetation layers, SNOWA (snow), and SOILA (soil) with 14 soil layers in the CMIP5 experiments.

\subsection{ISBA-CTRIP}
The land component for the CNRM-ESM2-1 model is presented by the Interaction Soil-Biosphere-Atmosphere (ISBA) land surface model and the total runoff integrating pathways (CTRIP) river routing model \citep[][]{Decharme2019, Delire2020}. ISBA-CTRIP simulates plant physiology, leaf phenology, carbon allocation and turnover, wild fires and carbon cycling through litter and soil  \citep[][]{Seferian2019}. Land use processes are prescribed instead of simulated, while land cover changes are used to represent anthropogenic disturbances. While the model does not include an interactive nitrogen cycle, its effects are included through an artificial downregulation of photosynthesis and a reduced specific leaf area with increasing CO$_2$ concentration. Six live and seven dead carbon pools are considered, along with 16 PFTs \citep[][]{Gibbelin2008}. Changes since the previous version used in CNRM-ESM-1 include improvements to the photosynthetic and autotrophic respiration schemes.

CTRIP includes carbon leaching through the soil and subsequent transport of dissolved organic carbon to the ocean. As chemical species such as dissolved inorganic carbon are not included, the air-water carbon exchange in the river routing model CTRIP cannot be computed. This leads to a carbon cycle which is not fully bounded.

\subsection{JSBACH}
JSBACH is the land component of the MPI-ESM model, with version 3.2 used for MPI-ESM1.2 \citep[][]{Mauritsen2019}. In the previous version, parameters in the model for decomposition of dead organic matter were tuned to reproduce the historical atmospheric CO$_2$ concentrations, with soil and litter carbon stocks merely being the result of this tuning. In version 3.2 decomposition is handled by the YASSO model \citep[][]{Yasso_Tuomi2011} based on litter and soil data resulting in no unconstrained parameters. YASSO simulates four fast soil carbon pools and one slow pool. A total of 18 dead carbon pools are considered due to a different application based on the woody and non-woody origins, as well as above and below ground decomposition. Additionally, three live carbon pools (natural vegetation, crops, pasture) and 13 PFTs are simulated by JSBACH, while permafrost carbon is not considered. The dynamical vegetation component interacts with the land use changes, modifying the land use data set to conform to the JSBACH setup. JSBACH3.2 includes an interactive terrestrial nitrogen cycle \citep[][]{Goll2017} driven by the nitrogen demand of the carbon cycle. Further adjustments in v3.2 include the change for carbon timescales in wood pools to be PFT specific.

\subsection{JULES}
The joint UK land environment simulator \citep[JULES,][]{JULES_Best2011, JULES_Clark2011} is the land model for the CMIP5 models HadGEM2-CC and HadGEM2-ES with the terrestrial carbon cycle following the Top-down Representation of Interaction of Foliage and Flora Including Dynamics \citep[TRIFFID, ][]{TRIFFID_Cox2001} dynamic vegetation scheme. The CMIP6 model UKESM1-0-LL \citep{Sellar2019} employs JULES version 5.0.

Improvements in the version used for UKESM1-0-LL include the introduction of nitrogen cycling, as well as developments to plant physiology and functional types and land use. In this model, nitrogen controls biomass and leaf area index within TRIFFID, thus only indirectly affecting photosynthetic capacity, as well as limiting the decomposition of litter into soil carbon. For better agreement with observations, global total GPP was tuned down through a reduction of the quantum efficiency of photosynthesis. Furthermore, crop and pasture areas were separated and a harvest carbon flux was introduced. UKESM1 has four soil carbon pools, nine natural PFTs - increased from five in prior versions - and four PFTs for crop and pasture.

\subsection{LM}
The GFDL Land Model \citep[LM,][]{LM2_Anderson2004} is used in the GFDL Earth System Models, with the CMIP5 models GFDL-ESM2G and GFDL-ESM2M \citep{Dunne2012, Dunne2013} using version 2.0 and the CMIP6 model GFDL-ESM4 employing version 4.1 \citep{Dunne2020}. Neither version includes an interactive nitrogen cycle.

Improvements since CMIP5 include updated soil types in the CORPSE model \citep{Sulman2014, Sulman2019}, hydrology, radiation, as well as the inclusion of a new fire model FINAL \citep{Rabin2018} with daily computations instead of previously annual figures and a new model for vegetation dynamics through the Perfect Plasticity Approximation \citep[PPA]{Weng2015}.
LM4.1 includes six live carbon pools for leaves, fine roots, heartwood, sapwood, seeds and nonstructural carbon, 20 vertical soil levels split into separate fast and slow pools and pools for soil microbes and microbial products. Six PFTs are included representing C3 grass, C4 grass, tropical trees, temperate deciduous trees, and cold evergreen trees. Land use is accounted for through annual wood harvesting, crop planting and harvesting, pasture grazing, and newly included rangelands.

\subsection{LPJ-GUESS}
The Lund-Potsdam-Jena General Ecosystem Simulator \citep[LPJ-GUESS,][]{Smith2014} in combination with the Hydrology Tiled ECMWF Scheme of Surface Exchanges over Land \citep[HTESSEL,][]{Balsamo2009} is the land model used in the EC-Earth models EC-Earth3-CC and EC-Earth3-Veg \citep[][]{Doescher2022}. The model described as "CC" additionally include ocean biogeochemistry (PISCES) and atmospheric composition for CO$_2$ (TM5), letting it perform CO$_2$ emission driven simulations.

HTESSEL solves the energy and water balance at the land surface, while vegetation types and vegetation coverage is interactively provided by the coupled LPJ-GUESS, which includes an interactive nitrogen cycle. Compared to the common area-based vegetation schemes, the interactive coupling of LPJ-GUESS to an atmospheric model should improve realism on longer timescales \citep{Doescher2022}. LPJ-GUESS includes 10 litter pools, seven vegetation carbon pools, as well as five soil carbon pools. Wildfires, disturbances and land use change are simulated on a yearly time step and distributed evenly throughout the year to conserve carbon mass. Land-use change dynamics are considered together with a crop module \citep{Lindeskog2013} including five crop functional types. Three types of plant phenology - evergreen, seasonal-deciduous, and stress-deciduous - are considered, with only the latter two being simulated with an explicit phenological cycle. Seasonal-deciduous PFTs  have a fixed growing season length of 210 days, while the growing season for stress-deciduous PFTs is determined by a threshold for the water stress.

\subsection{MATSIRO + SEIB-DGVM/VISIT-e}
The Minimal Advanced Treatments of Surface Interaction and RunOff \citep[MATSIRO,][]{Takata2003} is the physical land model for the MIROC-ESM family---MIROC-ESM and MIROC-ESM-CHEM (with coupled atmospheric chemistry) for CMIP5 \citep{Watanabe2011} and MIROC-ES2L for CMIP6 \citep{Hajima2020}---which consists of a single layer canopy, three snow layers and six soil layers down to a depth of 14m. For the CMIP6 version a physically based parametrization for snow distribution and snow-derived wetlands was added.
Biogeochemistry in MIROC-ESM and MIROC-ESM-CHEM is simulated by the Spatially Explicit Individual-Based Dynamic Global Vegetation Model \citep[SEIB-DGVM,][]{Sato2007}. It includes 13 PFTs split into two for grass and eleven for trees, as well as two organic carbon pools. Light capture competition among trees is explicitly modeled instead of parametrized.

MIROC-ES2L for CMIP6 uses the Vegetation integrative SImulator for Trace gases model \citep[VISIT,][]{Ito2012}, with changes for coupling to the ESM (adding the -e suffix), such as including leaf-nitrogen concentrations and thus limitations to enable fully coupled climate-carbon-nitrogen projections, and land-use change processes to get more use out of new LUC forcing data sets, such as using five types of land cover. The model does not simulate explicit dynamic vegetation. Three Vegetation carbon pools (leaf, stem, and root) are dynamically regulated and have constant turnover rates to three litter and three soil pools. 12 vegetation types are considered. A daily timestep is used for the land ecosystem and land biogeochemistry.

\subsection{ORCHIDEE}
The ORganizing Carbon and Hydrology in Dynamic EcosystEms \citep[ORCHIDEE,][]{Krinner2005, Boucher2020} land model is used in the IPSL models, version 1 for the CMIP5 models IPSL-CM5A-LR, IPSL-CM5A-MR, and version 2 in the CMIP6 model IPSL-CM5B-LR.
The model considers 15 PFTs, as well as 8 vegetation carbon, 4 litter carbon and 3 soil carbon pools. Plant and soil carbon fluxes are computed every 15\,min (the same as the atmospheric physics time step), while slow processes like soil and litter carbon dynamics are computed daily instead. The CMIP5 model used a two-layer bucket model for its soil hydrology, while in CMIP6 an 11-layer soil hydrology scheme is employed. Photosynthesis is parametrized based on the common Farquhar and Collatz schemes for C3 and C4, respectively. Nutrient limitation in CMIP6 is introduced through downregulation using a logarithmic function of the CO$_2$ concentration.

\subsection{Other: INMCM}
The carbon cycle module for INMCM \citep{Volodin2007} includes a single soil carbon pool. The most important changes with respect to INMCM4 for the CMIP6 models lie in the atmospheric component of the model, as well as some upgrades to the oceanic component, but no changes to the carbon cycle \citep{Volodin2017b}.

\noappendix       %% use this to mark the end of the appendix section. Otherwise the figures might be numbered incorrectly (e.g. 10 instead of 1).

%% Regarding figures and tables in appendices, the following two options are possible depending on your general handling of figures and tables in the manuscript environment:

%% Option 1: If you sorted all figures and tables into the sections of the text, please also sort the appendix figures and appendix tables into the respective appendix sections.
%% They will be correctly named automatically.

%% Option 2: If you put all figures after the reference list, please insert appendix tables and figures after the normal tables and figures.
%% To rename them correctly to A1, A2, etc., please add the following commands in front of them:

%\appendixfigures  %% needs to be added in front of appendix figures

%\appendixtables   %% needs to be added in front of appendix tables

%% Please add \clearpage between each table and/or figure. Further guidelines on figures and tables can be found below.

\authorcontribution{BG led the writing and analysis of the paper. MS assisted with parts of the programming. VE supervised the study. All authors contributed to the writing of the manuscript.} %% this section is mandatory

\competinginterests{At least one of the (co-)authors is a member of the editorial board of Biogeosciences.} %% this section is mandatory even if you declare that no competing interests are present

%\disclaimer{TEXT} %% optional section

\begin{acknowledgements}
This study was funded by the European Union’s Horizon 2020 research and innovation programmes "Climate-Carbon interactions in the Current Century" (4C; grant agreement No 821003) and "Earth System Models for the Future" (ESM2025; grant agreement No. 101003536). We acknowledge the World Climate Research Programme (WCRP), which, through its Working Group on Coupled Modelling, coordinated and promoted CMIP6. We thank the climate modeling groups (listed in Tables 2 and 3 of this paper) for producing and making available their model output, the Earth System Grid Federation (ESGF) for archiving the data and providing access, and the multiple funding agencies who support CMIP and ESGF. The computational resources of the Deutsches Klimarechenzentrum (DKRZ, Germany) were used to compute these results and are kindly acknowledged.

\end{acknowledgements}

%% REFERENCES

\bibliographystyle{copernicus}
\bibliography{gier2024}

%\printbibliography

%% The reference list is compiled as follows:

%\begin{thebibliography}{}

%\bibitem[AUTHOR(YEAR)]{LABEL1}
%REFERENCE 1

%\bibitem[AUTHOR(YEAR)]{LABEL2}
%REFERENCE 2

%\end{thebibliography}

%% Since the Copernicus LaTeX package includes the BibTeX style file copernicus.bst,
%% authors experienced with BibTeX only have to include the following two lines:
%%
%%\bibliographystyle{copernicus}
%%\bibliography{gier2022.bib}
%%
%% URLs and DOIs can be entered in your BibTeX file as:
%%
%% URL = {http://www.xyz.org/~jones/idx_g.htm}
%% DOI = {10.5194/xyz}

%% LITERATURE CITATIONS
%%
%% command                        & example result
%% \citet{jones90}|               & Jones et al. (1990)
%% \citep{jones90}|               & (Jones et al., 1990)
%% \citep{jones90,jones93}|       & (Jones et al., 1990, 1993)
%% \citep[p.~32]{jones90}|        & (Jones et al., 1990, p.~32)
%% \citep[e.g.,][]{jones90}|      & (e.g., Jones et al., 1990)
%% \citep[e.g.,][p.~32]{jones90}| & (e.g., Jones et al., 1990, p.~32)
%% \citeauthor{jones90}|          & Jones et al.
%% \citeyear{jones90}|            & 1990

%% FIGURES

% The figure files should be labelled correctly with Arabic numerals (e.g. fig01.jpg, fig02.png).

%
\clearpage

%t2
%\begin{sidewaystable*}[p]
%\begin{table*}[t]

% \scalebox{.80}[.80]{
%\begin{tabular}{lp{3cm}p{2.5cm}p{2.5cm}lp{3cm}}
\begin{longtable}{lp{3.0cm}p{2.7cm}p{2.5cm}lp{3cm}}
\caption{CMIP6 models analysed in this study. Under Comments, D stands for models including dynamic vegetation and N for models including Nitrogen cycles for models using land use emissions. Models for which emission driven simulations are also analysed are marked in bold script.}
\label{tab:CMIP6}\\
\tophline
{Model}&
{Institute}&
{Atmosphere model}&
{Land model}&
{Comment}&
{Main reference} \\
\hhline
\textbf{ACCESS-ESM1-5}&
Commonwealth Scientific and Industrial Research Organisation, Australia &
UM7.3 &
CABLE2.4, CASA-CNP &
N &
\citet{Law2017, Ziehn2017, Ziehn2020}\\
\hhline
\textbf{CanESM5}&
Canadian Center for Climate Modeling and Analysis, Canada&
CanAM5 &
CLASS3.6, CTEM1.2 &
D &
\citet{Swart2019}\\
\hhline
\textbf{CanESM5-CanOE}&
Canadian Center for Climate Modeling and Analysis, Canada&
CanAM5 &
CLASS3.6, CTEM1.2 &
D &
\citet{Swart2019}\\
\hhline
CESM2&
National Center for Atmospheric Research, USA&
CAM6&
CLM5&
N, D&
\citet{Danabasoglu2020}\\
\hhline
CESM2-WACCM&
National Center for Atmospheric Research, USA&
WACCM6&
CLM5&
N, D&
\citet{Danabasoglu2020}\\
\hhline
CMCC-CM2-SR5&
The Euro-Mediterranean Centre on Climate Change, Italy&
CAM5&
CLM4.5&
N&
\citet{Cherchi2019}\\
\hhline
CMCC-ESM2&
The Euro-Mediterranean Centre on Climate Change, Italy&
CAM5&
CLM4.5-BGC&
N&
\citet{Lovato2022}\\
\hhline
\textbf{CNRM-ESM2-1}&
CNRM-CERFACS, France&
ARPEGE-Climate v6.3 + SURFEX v8.0&
ISBA + CTRIP&
 &
\citet{Seferian2019}\\
\hhline
\textbf{EC-Earth3-CC}&
EC-Earth Consortium, Europe&
IFS 36r4 + HTESSEL + TM5&
LPJ-GUESS&
N, D&
\citet{Doescher2022}\\
\hhline
EC-Earth3-Veg&
EC-Earth Consortium, Europe&
IFS 36r4 + HTESSEL&
LPJ-GUESS&
N, D&
\citet{Doescher2022}\\
\hhline
\textbf{GFDL-ESM4}&
Geophysical Fluid Dynamics Laboratory, United States&
AM4.1&
LM4.1&
D&
\citet{Dunne2020}\\
\hhline
INM-CM4-8&
Institute for Numerical Mathematics, Russian Academy of Science, Russia&
Inbuilt&
Inbuilt&
 &
\citet{Volodin2018}\\
\hhline
INM-CM5-0&
Institute for Numerical Mathematics, Russian Academy of Science, Russia&
Inbuilt&
Inbuilt&
 &
\citet{Volodin2017a, Volodin2017b}\\
\hhline
IPSL-CM6A-LR&
L’Institut Pierre-Simon Laplace, France&
LMDZ6A&
ORCHIDEEv2&
 &
\citet{Boucher2020}\\
\hhline
\textbf{MIROC-ES2L}&
MIROC, Japan&
MIROC-AGCM + SPRINTARS&
VISIT-e \& MATSIRO6&
N&
\citet{Hajima2020}\\
\hhline
MPI-ESM-1-2-HAM&
HAMMOZ-Consortium, Europe&
ECHAM6.3– HAM2.3 &
JSBACH3.2&
 &
\citet{Neubauer2019, Tegen2019}\\
\hhline
\textbf{MPI-ESM1-2-LR}&
HAMMOZ-Consortium, Europe&
ECHAM6.3&
JSBACH3.2&
N, D &
\citet{Mauritsen2019}\\
\hhline
\textbf{MRI-ESM2-0}&
Meteorological Research Institute, Japan&
MRI-AGCM3.5 + MASINGAR mk-2r4c + MRI-CCM2.1&
HAL&
&
\citet{Yukimoto2019}\\
\hhline
\textbf{NorESM2-LM}&
NorESM Climate Modeling Consortium, Norway&
Modified CAM6&
CLM5&
N, D&
\citet{Seland2020}\\
\hhline
NorESM2-MM&
NorESM Climate Modeling Consortium, Norway&
Modified CAM6&
CLM5&
N, D&
\citet{Seland2020}\\
\hhline
\textbf{UKESM1-0-LL}&
Met Office Hadley Centre, United Kingdom&
Unified Model + UKCA&
JULES-ES-1.0&
N, D&
\citet{Sellar2019}\\
\hhline
SAM0-UNICON&
Seoul National University, Republic of Korea&
CAM5 + UNICON&
CLM4&
N&
\citet{Park2019}\\
\hhline
TaiESM1&
Research Centre for Environmental Changes, Academia Sinica, Taiwan&
Modified CAM5.3&
Modified CLM4&
N&
\citet{Lee2020}\\
\bottomhline
%\end{tabular}
\end{longtable}
%\end{table*}
%\end{sidewaystable*}

\clearpage

\begin{longtable}{lp{3.2cm}p{2.5cm}p{2.5cm}lp{3cm}}
\caption{CMIP5 models used in this study, notations as in Table 1.}
\label{tab:CMIP5}\\
\tophline
{Model}&
{Institute}&
{Atmosphere model}&
{Land model}&
{Comment}&
{Main reference} \\
\hhline
\textbf{BNU-ESM}&
College of Global Change and Earth System Science, China&
CAM3.5&
CoLM + BNU-DGVM&
D&
\citet{Ji2014}\\
\hhline
\textbf{CanESM2}&
Canadian Center for Climate Modeling and Analysis, BC, Canada&
CanAM4&
CLASS2.7 + CTEM1&
&
\citet{Arora2011}\\
\hhline
\textbf{CESM1-BGC}&
National Center for Atmospheric Research, United States&
CAM4&
CLM4&
N&
\citet{Hurrell2013}\\
\hhline
\textbf{GFDL-ESM2G}&
Geophysical Fluid Dynamics Laboratory, USA&
AM2&
LM3.0&
D&
\citet{Dunne2012, Dunne2013}\\
\hhline
\textbf{GFDL-ESM2M}&
Geophysical Fluid Dynamics Laboratory, USA&
AM2&
LM3.0&
D&
\citet{Dunne2012, Dunne2013}\\
\hhline
HadGEM2-CC&
Met Office Hadley Centre, United Kingdom&
Unified Model v6.6&
JULES + TRIFFID&
D&
\citet{Collins2011, HadGEM2011}\\
\hhline
HadGEM2-ES&
Met Office Hadley Centre, United Kingdom&
Unified Model v6.6&
JULES + TRIFFID&
D&
\citet{Collins2011, HadGEM2011}\\
\hhline
Inmcm4&
Institute for Numerical Mathematics, Russia&
Inbuilt&
Inbuilt&
&
\citet{Volodin2010}\\
\hhline
\textbf{FIO-ESM}&
The First Institute of Oceanography, SOA, China&
CAM3.0&
CLM3.5 + CASA&
&
\citet{Bao2012, Qiao2013}\\
\hhline
IPSL-CM5A-LR&
L'Institut Pierre-Simon Laplace, France&
LMDZ5&
ORCHIDEE&
&
\citet{Dufresne2013}\\
\hhline
IPSL-CM5B-LR&
L'Institut Pierre-Simon Laplace, France&
LMDZ5&
ORCHIDEE&
&
\citet{Dufresne2013}\\
\hhline
\textbf{MIROC-ESM}&
Japan Agency for Marine-Earth Science and Technology, Japan; Atmosphere and Ocean Research Institute, Japan&
MIROC-AGCM + SPRINTARS&
MATSIRO + SEIB-DGVM&
D&
\citet{Watanabe2011}\\
\hhline
MIROC-ESM-CHEM&
Japan Agency for Marine-Earth Science and Technology, Japan; Atmosphere and Ocean Research Institute, Japan&
MIROC-AGCM + SPRINTARS&
MATSIRO + SEIB-DGVM&
D&
\citet{Watanabe2011}\\
\hhline
\textbf{MPI-ESM-LR}&
Max Planck Institute for Meteorology, Germany&
ECHAM6&
JSBACH + BETHY&
D&
\citet{Giorgetta2013}\\
\hhline
MPI-ESM-MR&
Max Planck Institute for Meteorology, Germany&
ECHAM6&
JSBACH + BETHY&
D&
\citet{Giorgetta2013}\\
\hhline
\textbf{MRI-ESM1}&
Meteorological Research Institute, Japan&
MRI-AGCM3.3&
HAL&
&
\citet{Yukimoto2011}\\
\hhline
\textbf{NorESM1-ME}&
Norwegian Climate Center, Norway&
CAM4-Oslo&
CLM4&
N&
\citet{Tjiputra2013}\\
\bottomhline
%\end{tabular}
\end{longtable}
%\end{table*}
%\end{sidewaystable*}

\clearpage

%t3
\begin{table*}[t]
\caption{Reference data sets used in this study. data sets in bold are the main reference data set and those in italics the alternate reference for Figures \ref{fig:hist_perf}-\ref{fig:patterncor}.}
\label{tab:obs}
% \scalebox{.80}[.80]{
\begin{tabular}{p{3.2cm}p{3.2cm}p{3.2cm}p{2cm}p{3.3cm}}
\tophline

data set &
Source &
Variable &
Start Year &
Reference\\
\hhline
\textbf{JENA-CarboScope (sEXTocNEET\_v2020)} &
Inversion &
Land-Atmosphere Flux (NBP) &
1957 &
\citet{Roedenbeck2005}\\
\hhline
\textit{CAMS (v20r2)}&
Inversion &
Land-Atmosphere Flux (NBP) &
1979 &
\citet{Chevallier2005, Chevallier2010, Chevallier2013}\\
\hhline
GCP &
Dynamic global vegetation and bookkeeping model averages &
Land-Atmosphere Flux (NBP) &
1959 &
\citet{Friedlingstein2022, GCP2021}\\
\hhline
\textbf{FLUXCOM ANN-v1} &
Mix &
Gross Primary Productivity (GPP) &
1980 &
\citet{Jung2019}\\
\hhline
MTE &
Upscaled in situ &
Gross Primary Productivity (GPP) &
1982 &
\citet{Jung2011}\\
\hhline
\textit{GLASS} &
Satellite &
Gross Primary Productivity (GPP), Leaf Area Index (LAI) &
1982 (GPP), 1981 (LAI) &
\citet{Yuan2007_GLASS-GPP, Liang2021_GLASS}\\
\hhline
LAI3g &
Satellite &
Leaf Area Index (LAI) &
1981 &
\citet{Zhu2013}\\
\hhline
\textbf{LAI4g} &
Satellite &
Leaf Area Index (LAI) &
1982 &
\citet{Cao2023_lai4g}\\
\hhline
NDP-017b &
Mix &
Carbon Mass in Vegetation (cVeg) &
- &
\citet{Gibbs2006}\\
\hhline
HWSD+NCSCD &
Empirical &
Carbon Mass in Soil Pool (cSoil) &
- &
\citet{Wieder2014, Hugelius2013_NCSCD}\\
\bottomhline
\end{tabular}
\end{table*}

\clearpage

\clearpage

\clearpage

\clearpage

\clearpage

\clearpage

\clearpage

\clearpage

\clearpage

\clearpage

\clearpage

\clearpage

\clearpage

\clearpage

\clearpage

\clearpage

\clearpage

\clearpage

\clearpage

%%%%%%%%%%%%%%%%%% APPENDIX %%%%%%%%%%%%%%%%%
%% potentially move to supplementary material instead

%\appendixfigures  %% needs to be added in front of appendix figures

%\appendixtables

\newcommand{\beginsupplement}{%
    \setcounter{table}{0}
    \renewcommand{\thetable}{S\arabic{table}}%
    \setcounter{figure}{0}
    \renewcommand{\thefigure}{S\arabic{figure}}%
    \renewcommand\thesection{S\arabic{section}}
}

\beginsupplement
\section{Tables for Data from Scatterplots}

\input{tables/lai_mean_data}
\input{tables/lai_trend_data}

\input{tables/gpp_mean_data}
\input{tables/gpp_trend_data}

\input{tables/nbp_mean_data}
\input{tables/nbp_trend_data}

\input{tables/data_scatter_csoil_cveg}

\end{document}

%% file: tables/lai_mean_data.tex
\begin{longtable}{lllll}
\caption{LAI Mean for 1986-2005 [1]}\label{tab:lai_mean}\\
\toprule
Model &           Global &               NH &               SH &          Tropics \\
\midrule
\endfirsthead
\caption{LAI Mean for 1986-2005 [1]}\\
\toprule
Model &           Global &               NH &               SH &          Tropics \\
\midrule
\endhead
\midrule
\multicolumn{5}{r}{{Continued on next page}} \\
\midrule
\endfoot

\bottomrule
\endlastfoot
         BNU-ESM CMIP5e &  3.64 $\pm$ 0.08 &  2.13 $\pm$ 0.06 &   5.04 $\pm$ 0.1 &   6.37 $\pm$ 0.1 \\
         BNU-ESM CMIP5c &  3.75 $\pm$ 0.06 &   2.4 $\pm$ 0.09 &  4.98 $\pm$ 0.04 &  6.34 $\pm$ 0.06 \\
       CESM1-BGC CMIP5e &  2.68 $\pm$ 0.06 &  2.03 $\pm$ 0.06 &  3.31 $\pm$ 0.09 &   3.78 $\pm$ 0.1 \\
       CESM1-BGC CMIP5c &  2.64 $\pm$ 0.04 &  1.99 $\pm$ 0.05 &  3.27 $\pm$ 0.05 &  3.72 $\pm$ 0.05 \\
         CanESM2 CMIP5e &  1.71 $\pm$ 0.06 &  1.11 $\pm$ 0.02 &  2.26 $\pm$ 0.11 &  2.65 $\pm$ 0.13 \\
         CanESM2 CMIP5c &  1.72 $\pm$ 0.04 &  1.13 $\pm$ 0.02 &  2.26 $\pm$ 0.08 &   2.65 $\pm$ 0.1 \\
         FIO-ESM CMIP5e &  2.23 $\pm$ 0.01 &  1.69 $\pm$ 0.01 &   2.73 $\pm$ 0.0 &   3.14 $\pm$ 0.0 \\
      GFDL-ESM2G CMIP5e &   3.1 $\pm$ 0.06 &  2.65 $\pm$ 0.02 &   3.48 $\pm$ 0.1 &   4.09 $\pm$ 0.1 \\
\end{longtable}

%% file: tables/lai_trend_data.tex
\begin{longtable}{lllll}
\caption{LAI Trend for 1986-2005 [yr$^{-1}$]}\label{tab:lai_trend}\\
\toprule
                  Model &                Global &                   NH &                    SH &               Tropics \\
\midrule
\endfirsthead
\caption{LAI Trend for 1986-2005 [yr$^{-1}$]}\\
\toprule
                  Model &                Global &                   NH &                    SH &               Tropics \\
\midrule
\endhead
\midrule
\multicolumn{5}{r}{{Continued on next page}} \\
\midrule
\endfoot

\bottomrule
\endlastfoot
         BNU-ESM CMIP5e &    0.012 $\pm$ 0.0013 &   0.0072 $\pm$ 0.002 &    0.016 $\pm$ 0.0013 &   0.0158 $\pm$ 0.0015 \\
         BNU-ESM CMIP5c &   0.0102 $\pm$ 0.0008 &  0.0136 $\pm$ 0.0016 &   0.0057 $\pm$ 0.0012 &   0.0087 $\pm$ 0.0012 \\
       CESM1-BGC CMIP5e &    0.004 $\pm$ 0.0022 &   0.0067 $\pm$ 0.002 &   0.0011 $\pm$ 0.0034 &   0.0006 $\pm$ 0.0039 \\
       CESM1-BGC CMIP5c &   0.0035 $\pm$ 0.0016 &  0.0027 $\pm$ 0.0017 &    0.004 $\pm$ 0.0017 &   0.0057 $\pm$ 0.0016 \\
         CanESM2 CMIP5e &   0.0019 $\pm$ 0.0022 &   0.003 $\pm$ 0.0005 &   0.0005 $\pm$ 0.0043 &    0.0012 $\pm$ 0.005 \\
         CanESM2 CMIP5c &   0.0018 $\pm$ 0.0017 &   0.002 $\pm$ 0.0006 &   0.0012 $\pm$ 0.0033 &   0.0028 $\pm$ 0.0038 \\
         FIO-ESM CMIP5e &   0.0005 $\pm$ 0.0002 &  0.0009 $\pm$ 0.0003 &  -0.0001 $\pm$ 0.0001 &  -0.0001 $\pm$ 0.0001 \\
      GFDL-ESM2G CMIP5e &   0.0055 $\pm$ 0.0019 &  0.0028 $\pm$ 0.0008 &   0.0077 $\pm$ 0.0035 &    0.008 $\pm$ 0.0035 \\
\end{longtable}

%% file: tables/gpp_mean_data.tex
\begin{longtable}{lllll}
\caption{GPP Mean for 1986-2005 [PgC yr$^{-1}$]}\label{tab:gpp_mean}\\
\toprule
                  Model &             Global &                NH &                 SH &            Tropics \\
\midrule
\endfirsthead
\caption{GPP Mean for 1986-2005 [PgC yr$^{-1}$]}\\
\toprule
                  Model &             Global &                NH &                 SH &            Tropics \\
\midrule
\endhead
\midrule
\multicolumn{5}{r}{{Continued on next page}} \\
\midrule
\endfoot

\bottomrule
\endlastfoot
         BNU-ESM CMIP5e &   83.54 $\pm$ 2.48 &   30.35 $\pm$ 1.3 &   53.19 $\pm$ 1.54 &   47.13 $\pm$ 1.18 \\
         BNU-ESM CMIP5c &    86.26 $\pm$ 2.0 &  33.66 $\pm$ 1.35 &    52.6 $\pm$ 0.92 &   46.93 $\pm$ 0.92 \\
       CESM1-BGC CMIP5e &  107.42 $\pm$ 2.23 &   34.94 $\pm$ 1.1 &    72.49 $\pm$ 1.3 &   61.98 $\pm$ 1.12 \\
       CESM1-BGC CMIP5c &   106.12 $\pm$ 1.7 &   34.5 $\pm$ 0.78 &   71.62 $\pm$ 1.06 &   61.24 $\pm$ 0.93 \\
         CanESM2 CMIP5e &  103.74 $\pm$ 2.72 &  32.32 $\pm$ 0.86 &   71.42 $\pm$ 2.76 &   56.61 $\pm$ 2.07 \\
         CanESM2 CMIP5c &  104.37 $\pm$ 2.41 &  32.66 $\pm$ 0.74 &   71.71 $\pm$ 2.09 &   56.66 $\pm$ 1.78 \\
         FIO-ESM CMIP5e &   151.2 $\pm$ 2.28 &  44.86 $\pm$ 0.99 &  106.34 $\pm$ 1.69 &   91.94 $\pm$ 1.59 \\
      GFDL-ESM2G CMIP5e &  139.01 $\pm$ 4.89 &  46.29 $\pm$ 1.23 &   92.72 $\pm$ 4.42 &   75.82 $\pm$ 2.78 \\
\end{longtable}

%% file: tables/gpp_trend_data.tex
\begin{longtable}{lllll}
\caption{GPP Trend for 1986-2005 [PgC yr$^{-2}$]}\label{tab:gpp_trend}\\
\toprule
                  Model &               Global &                   NH &                    SH &               Tropics \\
\midrule
\endfirsthead
\caption{GPP Trend for 1986-2005 [PgC yr$^{-2}$]}\\
\toprule
                  Model &               Global &                   NH &                    SH &               Tropics \\
\midrule
\endhead
\midrule
\multicolumn{5}{r}{{Continued on next page}} \\
\midrule
\endfoot

\bottomrule
\endlastfoot
         BNU-ESM CMIP5e &  0.3915 $\pm$ 0.0353 &  0.1594 $\pm$ 0.0357 &   0.2321 $\pm$ 0.0283 &    0.1665 $\pm$ 0.026 \\
         BNU-ESM CMIP5c &  0.3177 $\pm$ 0.0272 &   0.2067 $\pm$ 0.023 &    0.111 $\pm$ 0.0258 &   0.1123 $\pm$ 0.0257 \\
       CESM1-BGC CMIP5e &  0.2964 $\pm$ 0.0552 &  0.1536 $\pm$ 0.0246 &   0.1428 $\pm$ 0.0395 &   0.1127 $\pm$ 0.0359 \\
       CESM1-BGC CMIP5c &  0.2269 $\pm$ 0.0419 &  0.0828 $\pm$ 0.0242 &   0.1441 $\pm$ 0.0254 &     0.1322 $\pm$ 0.02 \\
         CanESM2 CMIP5e &  0.1535 $\pm$ 0.1023 &   0.113 $\pm$ 0.0215 &   0.0405 $\pm$ 0.1098 &   0.0422 $\pm$ 0.0818 \\
         CanESM2 CMIP5c &  0.1575 $\pm$ 0.0886 &  0.0826 $\pm$ 0.0222 &    0.075 $\pm$ 0.0812 &   0.0879 $\pm$ 0.0681 \\
         FIO-ESM CMIP5e &  0.3111 $\pm$ 0.0536 &  0.1364 $\pm$ 0.0228 &   0.1746 $\pm$ 0.0536 &   0.1641 $\pm$ 0.0502 \\
      GFDL-ESM2G CMIP5e &  0.4753 $\pm$ 0.1595 &   0.161 $\pm$ 0.0312 &   0.3143 $\pm$ 0.1598 &   0.2258 $\pm$ 0.0971 \\
\end{longtable}

%% file: tables/nbp_mean_data.tex
\begin{longtable}{lllll}
\caption{NBP Mean for 1986-2005 [PgC yr$^{-1}$]}\label{tab:nbp_mean}\\
\toprule
                  Model &            Global &                NH &                SH &           Tropics \\
\midrule
\endfirsthead
\caption{NBP Mean for 1986-2005 [PgC yr$^{-1}$]}\label{tab:nbp_mean}\\
\toprule
                  Model &            Global &                NH &                SH &           Tropics \\
\midrule
\endhead
\midrule
\multicolumn{5}{r}{{Continued on next page}} \\
\midrule
\endfoot

\bottomrule
\endlastfoot
         BNU-ESM CMIP5e &   0.36 $\pm$ 0.92 &   0.39 $\pm$ 0.69 &  -0.03 $\pm$ 0.94 &   0.01 $\pm$ 0.96 \\
         BNU-ESM CMIP5c &   1.32 $\pm$ 1.15 &   0.89 $\pm$ 0.59 &   0.43 $\pm$ 1.04 &   0.39 $\pm$ 0.99 \\
       CESM1-BGC CMIP5e &  -0.08 $\pm$ 0.97 &   0.06 $\pm$ 0.52 &  -0.14 $\pm$ 0.65 &  -0.11 $\pm$ 0.55 \\
       CESM1-BGC CMIP5c &  -0.14 $\pm$ 0.67 &   0.02 $\pm$ 0.36 &   -0.17 $\pm$ 0.6 &  -0.12 $\pm$ 0.41 \\
         CanESM2 CMIP5e &    0.3 $\pm$ 2.47 &   0.23 $\pm$ 0.54 &   0.08 $\pm$ 2.59 &   0.12 $\pm$ 1.94 \\
         CanESM2 CMIP5c &   0.21 $\pm$ 2.07 &   0.11 $\pm$ 0.34 &    0.1 $\pm$ 1.89 &    0.07 $\pm$ 1.4 \\
         FIO-ESM CMIP5e &   1.84 $\pm$ 0.64 &   0.78 $\pm$ 0.17 &   1.06 $\pm$ 0.61 &   0.91 $\pm$ 0.62 \\
      GFDL-ESM2G CMIP5e &    0.9 $\pm$ 2.95 &   0.39 $\pm$ 0.78 &    0.5 $\pm$ 2.83 &   0.23 $\pm$ 2.01 \\
\end{longtable}

%% file: tables/nbp_trend_data.tex
\begin{longtable}{lllll}
\caption{NBP Trend for 1986-2005 [PgC yr$^{-2}$]}\label{tab:nbp_trend}\\
\toprule
                  Model &                Global &                    NH &                    SH &               Tropics \\
\midrule
\endfirsthead
\caption{NBP Trend for 1986-2005 [PgC yr$^{-2}$]}\\
\toprule
                  Model &                Global &                    NH &                    SH &               Tropics \\
\midrule
\endhead
\midrule
\multicolumn{5}{r}{{Continued on next page}} \\
\midrule
\endfoot

\bottomrule
\endlastfoot
         BNU-ESM CMIP5e &   0.0343 $\pm$ 0.0358 &   0.0217 $\pm$ 0.0269 &   0.0127 $\pm$ 0.0373 &  -0.0003 $\pm$ 0.0385 \\
         BNU-ESM CMIP5c &  -0.0049 $\pm$ 0.0457 &  -0.0067 $\pm$ 0.0235 &   0.0018 $\pm$ 0.0415 &   0.0087 $\pm$ 0.0396 \\
       CESM1-BGC CMIP5e &    0.0468 $\pm$ 0.037 &   0.0331 $\pm$ 0.0192 &   0.0137 $\pm$ 0.0256 &   0.0117 $\pm$ 0.0219 \\
       CESM1-BGC CMIP5c &   0.0251 $\pm$ 0.0259 &   0.0255 $\pm$ 0.0131 &   -0.0003 $\pm$ 0.024 &   0.0037 $\pm$ 0.0164 \\
         CanESM2 CMIP5e &  -0.0427 $\pm$ 0.0979 &   0.0048 $\pm$ 0.0216 &  -0.0475 $\pm$ 0.1028 &  -0.0404 $\pm$ 0.0767 \\
         CanESM2 CMIP5c &  -0.0579 $\pm$ 0.0815 &  -0.0105 $\pm$ 0.0135 &  -0.0473 $\pm$ 0.0746 &  -0.0264 $\pm$ 0.0555 \\
         FIO-ESM CMIP5e &   0.0323 $\pm$ 0.0243 &    0.0128 $\pm$ 0.006 &    0.0195 $\pm$ 0.024 &   0.0218 $\pm$ 0.0242 \\
      GFDL-ESM2G CMIP5e &   0.0966 $\pm$ 0.1154 &   0.0597 $\pm$ 0.0276 &    0.037 $\pm$ 0.1126 &   0.0378 $\pm$ 0.0796 \\
\end{longtable}

%% file: tables/data_scatter_csoil_cveg.tex
\begin{longtable}{lrr}
\caption{Carbon Storage Totals [PgC]} \label{tab:csoil-cveg} \\
\toprule
 & cVeg & cSoil + cLitter \\
\midrule
\endfirsthead
\caption[]{Carbon Storage Totals [PgC]} \\
\toprule
 & cVeg & cSoil + cLitter \\
\midrule
\endhead
\midrule
\multicolumn{3}{r}{Continued on next page} \\
\midrule
\endfoot
\bottomrule
\endlastfoot
BNU-ESM CMIP5e & 1205 & 856 \\
BNU-ESM CMIP5c & 1259 & 913 \\
CESM1-BGC CMIP5e & 538 & 515 \\
CESM1-BGC CMIP5c & 532 & 513 \\
GFDL-ESM2G CMIP5e & 680 & 1434 \\
GFDL-ESM2G CMIP5c & 660 & 1422 \\
GFDL-ESM2M CMIP5e & 802 & 1292 \\
GFDL-ESM2M CMIP5c & 766 & 1279 \\
\end{longtable}